\documentclass[aps,showpacs]{revtex4}
\usepackage[dvips]{epsfig,color}
\include{graphics}
 
\usepackage{graphicx}
\usepackage{dcolumn}

\begin{document}
\title{New fit to the reaction $\gamma p\rightarrow K^+\Lambda$}
\author{Alejandro de la Puente, Oren V. Maxwell, and Brian A. Raue}
\affiliation{Department of Physics, Florida International University,
University Park, Miami, Florida 33199, USA}

{\centerline{\today{}}}
\begin{abstract}

The reaction $\gamma p\rightarrow K^+\Lambda$ has been investigated
over the center-of-momentum energy, $W$, range from threshold up to
2.2 GeV in a tree-level effective Lagrangian model that incorporates
most of the well-established baryon resonances with spins equal to or
below $\frac{5}{2}$.  Four less well-established nucleon resonances of
higher mass are also included. The fitted parameters consist, for each
resonance included, of the products of the coupling strengths at the
electromagnetic and strong interaction vertices and, for the
less-established nucleon resonances, the total decay width. For the
well-established nucleon resonances, the energy and momentum
dependence of the widths is treated within a dynamical model that is
normalized to give the empirical decay branching ratios on the
resonance mass shells.  For the less-established resonances, the total
decay width is treated as a single parameter independent of the
reaction kinematics. The model is used to fit recent data for the
unpolarized differential cross section (CLAS), the induced hyperon
polarization asymmetry, $P$ (CLAS, GRAAL, and SAPHIR), the beam spin
asymmetry, $\Sigma$ (LEPS), and the double
polarization observables $C_x$ and $C_z$ (CLAS). Two different fits
were obtained: one that incorporates SU(3) symmetry constraints on
the Born contributions to the reaction amplitude and one in which
these constraints are relaxed. Explicit numerical results are given
only for the first fit since the two fits gave nearly identical
results for the observables and the $\chi^2$ per degree of freedom
obtained with the second fit was only marginally better than that of
the first fit ($<1\%$ better). Results are presented for the fitted
observables at several different energies and center-of-momentum
(c.m.) frame kaon angles.

\end{abstract}

\pacs{25.10.+s, 25.20.Lj, 13.60.-r} 
\maketitle
\section{introduction}

Interest in the electromagnetic production of strangeness from few nucleon 
targets, such as the proton and the deuteron, dates back to the 1960's 
\cite{thom,renard,early}, but it is comparatively recently that high quality 
data, suitable for quantitatively testing theoretical models, has become
available
\cite{saphir,leps,hallc,claspol,classig,graal,newsig,newpol,clasep,clasltp}. 
The strangeness degree of freedom imparts to these reactions the potential to 
provide fundamental information concerning both the strong and electromagnetic
interactions beyond that obtainable from reactions involving just the
non-strange baryons.  Within an effective Lagrangian model, one could use
these reactions to search for baryon resonances that decay to strange
particles and possibly test SU(3) symmetry relations among the couplings of
resonances within the same SU(3) multiplets.  By comparing photoproduction and
electroproduction results, one might be able to extract information concerning
the electromagnetic form factors of baryon resonances.  Finally, with the
aid of a quantitative model for electromagnetic strangeness production from
the proton, one could, within the impulse approximation, use results for
strangeness production from the deuteron and other light nuclei to study final
state interactions involving the $\Lambda$ and $\Sigma$ baryons.

Much of the theoretical work over the past 20 years or so has been based
on effective Lagrangian models \cite{abw,aws,wjc,mbh,mb,korea,sac1,
sac2,hly,ctls,ghent,max1,max2,max3,us}.  Recently, there have been
several coupled channel analyses
\cite{bonn,kaiser,feuster,jdiaz} that have revealed the need for
resonances that had not been previously included in many of the
effective Lagrangian models.  Until recently, the fits and models
were largely based on older data, and often combined photoproduction
data and electroproduction data to generate the fits.  More recent
fits have made use of various combinations of recent data from the
SAPHIR \cite{saphir}, CLAS \cite{classig,newsig,newpol}, LEPS
\cite{leps}, and GRAAL \cite{graal} collaborations.

In Ref.~\cite{max3}, it was suggested that the photoproduction data and the
electroproduction data should {\em not} be fit together; rather one should
first generate a model for the basic reaction using photoproduction data
alone and then use that model, in conjunction with electroproduction data,
to obtain information concerning the electromagnetic form factors of the
various resonances in the model.  This consideration, along with the
abundance of new data, particularly, polarization data, has motivated us to
develop a new model for the reaction $\gamma p\rightarrow K^+\Lambda$ over
the energy range from threshold up to a center-of-momentum (c.m.) energy of
2.2 GeV.  Although the CLAS data extends up to 2.6 GeV, the lack of
$s$-channel resonances in our model with masses above 2.2 GeV precludes a
reliable treatment of the higher energy data.  As discussed in
Sec.~\ref{sec-results}, the fits involve a subtle interference between
$s$-channel contributions to the reaction on the one hand and $u$ and
$t$-channel contributions on the other hand, which, to be effective,
requires that the $s$-channel resonances included in a particular fit have
masses spanning the whole energy range of that fit.

The model is similar to that described in Refs.~\cite{max1,max2,max3}, but has
been expanded to include spin $\frac{5}{2}$ baryon resonances in both the $s$
and $u$-channels, in addition to the spin $\frac{1}{2}$ and spin $\frac{3}{2}$ 
resonances included in the earlier work.  The present model also includes 
several higher energy, less well-established resonances in the $s$-channel that
were not included in the earlier work.  Finally, the fits described here are 
much more elaborate than those described in Refs.~\cite{max1,max2,max3} in 
that both single and double polarization data are included in the present fits.
The reaction model is described in some detail in Sec.~\ref{sec-model}.

Two separate fits were generated.  The first fit has SU(3) symmetry constraints
imposed on the Born terms in all three channels. These constraints, along with 
inputted empirical values for the baryon magnetic moments, require particular 
relationships between the various Born terms and also, in conjunction with 
other considerations, provide a range of values for the $\Lambda K N$ coupling.
This is discussed more fully in Sec.~\ref{sec-fitting}. In the second fit, the
$\Lambda K N$ coupling was allowed to move outside the SU(3) symmetry range and
assume whatever value yielded the best fit to the data. The second fit obtained
is nearly identical to the first fit and yielded a $\chi^2$ per degree of 
freedom that differs from that of the first fit by less than $1\%$.  For these
reasons, only results derived from the first fit are presented here.  

We have used all recently published results for the spin observables,
$P$ \cite{classig,saphir,graal}, $\Sigma$ \cite{leps,graal}, $C_x$ and $C_z$
\cite{newpol} in our fits. To avoid difficulties associated with
inconsistencies between different data sets \cite{bydzovsky},
only the most recent cross section data from the CLAS collaboration
\cite{newsig} were used in the fits.  Further details concerning the
fitting procedure are contained in Sec.~\ref{sec-fitting}. 

The resulting fit, along with several figures illustrating the quality of the
fits, are presented and discussed in Sec.~\ref{sec-results}.
Sec.~\ref{sec-results} also contains some concluding remarks and a brief
discussion of future work.

\section{the reaction model}
\label{sec-model}

\begin{figure}[tbph]
\vspace{5.2cm}
\includegraphics{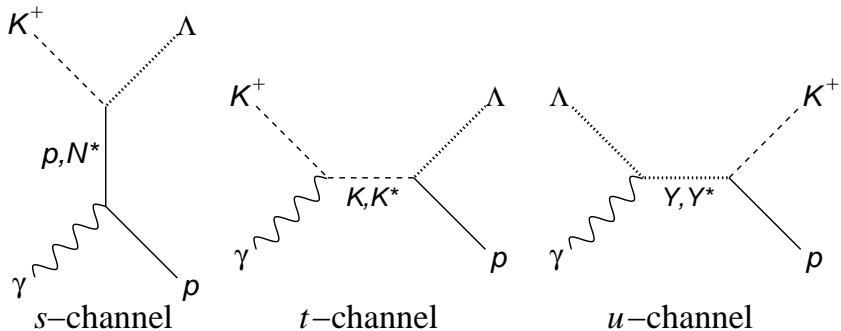}
\caption{Contributions to the amplitude for the reaction $\gamma p 
\rightarrow K^+\Lambda$.}
\label{fig-chan}
\end{figure}

The reaction model incorporates contributions in the $s$-channel, $u$-channel,
and $t$-channel.  These are illustrated in Fig.~\ref{fig-chan}. The
$s$-channel contributions include the Born term with an intermediate proton
and contributions in which an intermediate nucleon resonance is excited.
Similarly, the $u$-channel Born contributions with the excitation of an
intermediate $\Lambda$ or $\Sigma$ baryon are supplemented by contributions
involving the excitation of an intermediate hyperon resonance. In the
$t$-channel, contributions from both $K^{\star}(892)$ and $K1(1270)$ exchange
are included, as well as the Born contribution involving ground state kaon
exchange.

In the various channels, the reaction amplitudes have the general forms
\begin{equation}
\hat{T}_s = \sum_{N^{\star}}{\cal V}^{\dagger}_K(p_K) D(p_s){\cal V}_{\gamma} 
      (p_{\gamma})  \label{schan} 
\end{equation}
\begin{equation}
\hat{T}_u = \sum_{Y^{\star}}{\cal V}^{\dagger}_{\gamma}(p_{\gamma}) 
    D(p_u) {\cal V}_K(p_K), \label{uchan} 
\end{equation}
and
\begin{equation}
\hat{T}_t = \sum_{K^{\star}}{\cal V}^{\dagger}_{\gamma K}
(p_{\gamma},p_t) D_t(p_t) {\cal V}_{p\Lambda}(p_t), \label{tchan} 
\end{equation}
where $p_s=p_{\Lambda}+p_K$, $p_u=p_{\Lambda}-p_{\gamma}$, and 
$p_t=p_{\gamma}-p_K$ are the intermediate 4-momenta in the various channels,  
the $\cal{V}$'s are the electromagnetic and strong interaction vertices, and 
the $D$'s are the associated intermediate baryon and meson propagators.
In all channels the forms of both the electromagnetic vertices and the strong 
interaction vertices depend on the spin and the parity of the intermediate 
hadron that is excited.

In the $t$-channel, the two vertices are given by 
\begin{equation}
{\cal V}_{\gamma K} = -e \epsilon \cdot (p_K-p_t) \label{tphk} 
\end{equation}
and
\begin{equation}
{\cal V}_{p\Lambda} = g_{\Lambda Kp} \gamma_5             \label{tlamvk}
\end{equation}
for an intermediate ground state kaon (the $t$-channel Born term), by
\begin{equation}
{\cal V}^{\mu}_{\gamma K} = \frac{g_{\gamma K K^{\star}}}{m_{sc}} 
      \epsilon^{\mu\nu\rho\lambda} \epsilon_{\nu}p_{\gamma\rho} p_{t \lambda}
      \label{tphk*} 
\end{equation}
and
\begin{equation}
{\cal V}^{\mu}_{p\Lambda} = (g^V_{\Lambda K^{\star}p}+
\frac{g^T_{\Lambda K^{\star}p}}{m_p+m_{\Lambda}}\gamma\cdot p_t)\gamma^{\mu}
      \label{tlamvk*}
\end{equation}
for an intermediate $K^{\star}(892)$ resonance and by
\begin{equation}
{\cal V}^{\mu}_{\gamma K} = \frac{g_{\gamma K K1}}{m_{sc}} 
      (\epsilon \cdot p_t p^{\mu}_{\gamma}-p_{\gamma} \cdot p_t \epsilon^{\mu})
      \label{tphk1} 
\end{equation}
and
\begin{equation}
{\cal V}^{\mu}_{p\Lambda} = (g^V_{\Lambda K1 p}+\frac{g^T_{\Lambda K1 p}}
      {m_p+m_{\Lambda}}\gamma\cdot p_t)\gamma^{\mu}\gamma_5  \label{tlamvk1}
\end{equation}
for an intermediate $K1(1270)$ resonance. Here, $\epsilon$ is the photon 
polarization 4-vector and $m_{sc}$ is a scaling mass, set equal to 
1000 MeV, that is introduced to make the electromagnetic coupling strengths 
dimensionless.  The corresponding kaon resonance propagators both have
the same form:
\begin{equation}
D_t = \frac{-g_{\mu\nu}+\frac{p_{t\mu}p_{t\nu}}{m_{K^{\star}}^2}}
{p_t^2-m_{K^{\star}}^2},    \label{kprop}
\end{equation}
where the label $K^{\star}$ now refers to either of the two resonances. 
We note that the propagator employed here, in contrast with that used in 
Refs.~\cite{max1,max2,max3}, does not contain a width since the intermediate 
energies in the $t$-channel lie well below the thresholds of any possible decay
channels.

In the $s$ and $u$-channel Born contributions and in those contributions 
arising from the excitation of intermediate spin $\frac{1}{2}$ resonances, we 
employ standard expressions for the electromagnetic vertices and use the 
pseudoscalar form for the strong interaction vertices. For positive parity 
intermediate baryons, this gives
\begin{equation}
{\cal V}_{K \frac{1}{2}^+}(p_K) = g \gamma_5  \label{K12} 
\end{equation} 
and
\begin{equation}
{\cal V}_{\gamma \frac{1}{2}^+}(p_{\gamma}) = g_{\gamma} 
      \epsilon_{\mu} i \sigma^{\mu\nu}(p_{\gamma})_{\nu}  \label{phv12} 
\end{equation}
with 
\begin{equation}
g_{\gamma} = \frac{e\kappa}{2m_B},  \label{phcoup}
\end{equation}
where $\kappa$ is defined by its relation to the transition magnetic moment,
\begin{equation}
\mu_T = \frac{e\kappa}{m_B+m_I}, \label{mut}
\end{equation}
$m_B$ is the mass of the incoming or outgoing baryon ($m_p$ or $m_{\Lambda}$), 
and $m_I$ is the mass of the intermediate baryon. The corresponding 
expressions for negative parity intermediate baryons just have the
$\gamma_5$ factor transposed from the strong interaction vertex to the 
electromagnetic vertex.  For intermediate protons, there is an additional 
term,
\begin{equation}
{\cal V}_{charge}(p_{\gamma}) = e\gamma^{\mu} \epsilon_{\mu},  \label{charge}
\end{equation}
arising from the proton's charge.  For the spin $\frac{1}{2}$ propagator, we
employ, in agreement with other authors, a relativistic Breit-Wigner form, 
\begin{equation}
D^{\frac{1}{2}}(p) = \frac{\gamma\cdot p+m_I}{p^2-m^2_I+im_I \Gamma_I},
     \label{rprop12}
\end{equation} 
where the width $\Gamma_I$ is non-zero only in the $s$-channel resonance 
contributions.

In the $s$ and $u$-channel contributions from intermediate spin $\frac{3}{2}$
baryons, a number of different forms have been employed for the vertices and 
the propagator.  The authors of Ref.~\cite{abw} introduced a form for the 
spin $\frac{3}{2}$ propagator in the $s$-channel in which the intermediate 
baryon mass appearing in the numerator and projection operator of the 
Rarita-Schwinger propagator was replaced by $\sqrt{s}$. This was motivated by 
the desire to ensure gauge invariance off-shell.  However, use of the same 
prescription in the $u$-channel leads to unphysical singularities.  Moreover, 
as pointed out in Ref.~\cite{muk}, the propagator employed in Ref.~\cite{abw}
does not satisfy the differential equation that defines the propagator as a 
Green's function.  For these reasons, we employ the standard Rarita-Schwinger 
form for the spin $\frac{3}{2}$ propagator and use forms for the corresponding
vertices that are similar to those introduced in Ref.~\cite{sac2}. However,
in contrast to the work of Ref.~\cite{sac2}, we make no attempt to include 
off-shell terms in the vertices. The results reported in Ref.~\cite{sac2} 
suggest that these terms, as well as off-shell terms in the propagator, 
have a relatively modest effect on the calculated observables.  The resulting
vertices for positive parity spin $\frac{3}{2}$ intermediate baryons take the 
forms 
\begin{equation}
{\cal V}_{K \frac{3}{2}^+}^{\mu}(p_K) = -\frac{g}{m_{\pi}} p_K^{\mu}, 
\label{K32} 
\end{equation} 
and
\begin{equation}
{\cal V}_{\gamma \frac{3}{2}^+}^{\mu}(p_{\gamma}) = [\frac{g_1}{2m_B}
 (\epsilon^{\mu}\gamma\cdot p_{\gamma}-p_{\gamma}^{\mu}\gamma\cdot\epsilon)
      +\frac{g_2}{4m_B^2}(\epsilon\cdot p_B p_{\gamma}^{\mu}
      - p_{\gamma}\cdot p_B \epsilon^{\mu})] \gamma_5, \label{phv32}
\end{equation}
where $p_B$ is the 4-momentum of the incoming or outgoing ground state baryon.
As for the spin $\frac{1}{2}$ contributions, the negative parity vertices just 
have the $\gamma_5$ factor transposed from one vertex to the other vertex.
Note in these expressions that the strong interaction coupling has been 
divided by the pion mass, rather than the kaon mass as in Ref.~\cite{sac2}.  
This makes it easier to compare our coupling strengths with the corresponding 
pion couplings for the purpose of testing SU(3) symmetry relations among the
couplings. The Rarita-Schwinger propagator is obtained by multiplying the spin 
$\frac{1}{2}$ propagator given by Eq.~(\ref{rprop12}) on the right by the 
spin $\frac{3}{2}$ projection operator
\begin{equation}
P^{\frac{3}{2}}_{\mu\nu} = g_{\mu\nu}-\frac{1}{3}\gamma_{\mu}\gamma_{\nu}
    +\frac{1}{3}\frac{p_{\mu}\gamma_{\nu}-p_{\nu}\gamma_{\mu}}{m_I}
    -\frac{2}{3}\frac{p_{\mu}p_{\nu}}{m_I^2}.     \label{P32}
\end{equation} 

With the exception of Ref.~\cite{sac1}, most of the earlier work on the
photoproduction of strangeness does not include contributions from
intermediate states with spin $\frac{5}{2}$, even though there are several
well-established baryon resonances with this spin below 2 GeV.
Ref.~\cite{sac1} and some of the later work do include spin $\frac{5}{2}$
resonances in the $s$-channel but not in the $u$-channel.  To our knowledge,
the present work is the first analysis to include resonances with spin greater
than $\frac{3}{2}$ in {\em both} the $s$ and $u$-channels.  For the spin
$\frac{5}{2}$ vertices, we employ forms similar to those given in
Ref.~\cite{sac1} but modified so as to be consistent with the forms adopted
here for the spin $\frac{3}{2}$ vertices. Again, we do not include any
off-shell terms in the spin $\frac{5}{2}$ vertices.  For positive parity
intermediate resonances, the resulting vertices are given by
\begin{equation}
{\cal V}_{K \frac{5}{2}^+}^{\mu\nu}(p_K) = \frac{g}{m_{\pi}^2} p_K^{\mu} 
        p_K^{\nu} \gamma_5        \label{K52} 
\end{equation} 
and
\begin{equation}
{\cal V}_{\gamma \frac{5}{2}^+}^{\mu\nu}(p_{\gamma}) = [\frac{g_1}{2m_B}
 (\epsilon^{\mu}\gamma\cdot p_{\gamma}-p_{\gamma}^{\mu}\gamma\cdot\epsilon)
      +\frac{g_2}{4m_B^2}(\epsilon\cdot p_B p_{\gamma}^{\mu}
      - p_{\gamma}\cdot p_B \epsilon^{\mu})] \frac{p_{\gamma}^{\nu}}{m_{\pi}}.
                \label{phv52}
\end{equation}
As for the other $s$ and $u$-channel vertices, the corresponding negative 
parity vertices just have the $\gamma_5$ factor transposed from one vertex to 
the other vertex.  The corresponding propagator is constructed by multiplying 
the spin $\frac{1}{2}$ propagator on the right by the spin $\frac{5}{2}$ 
projector operator, 
\begin{eqnarray}
P^{\frac{5}{2}}_{\mu\nu,\mu^{\prime}\nu^{\prime}} &=& 
     R_{\mu\nu,\mu^{\prime}\nu^{\prime}} 
   -\frac{1}{5}P_{\mu\nu}P_{\mu^{\prime}\nu^{\prime}}-\frac{1}{5}
(P_{\mu\rho}\gamma^{\rho}\gamma^{\sigma}R_{\sigma\nu,\mu^{\prime}\nu^{\prime}}
+P_{\nu\rho}\gamma^{\rho}\gamma^{\sigma}R_{\sigma\mu,\mu^{\prime}\nu^{\prime}})
\end{eqnarray}                   \label{P52}
with
\begin{equation}
R_{\mu\nu,\mu^{\prime}\nu^{\prime}} =  
        \frac{1}{2}(P_{\mu\mu^{\prime}}P_{\nu\nu^{\prime}}
        +P_{\mu\nu^{\prime}}P_{\nu\mu^{\prime}}),           \label{Rdef}
\end{equation}  
where 
\begin{equation}
P_{\mu\nu} = g_{\mu\nu}-p_{\mu}p_{\nu}/m_I^2.        \label{Pdef}
\end{equation}

\subsection{$s$-channel resonance widths}

The intermediate nucleon resonances excited in the $s$-channel generally lie 
at energies above the thresholds for decay into various decay channels.  Thus, 
the propagators employed for the $s$-channel resonances need to include widths,
and these widths are generally required rather far off the resonance mass 
shells.  Most previous studies have ignored the off-shell nature
of the resonances and simply used the on-shell values of the widths.  In 
Ref.~\cite{max1} a model was proposed to dynamically generate widths off-shell
by making use of partial width data summarized in the particle data tables
\cite{pdt}.  The full width is first decomposed into a number of different 
decay channels.  In each such channel, the off-shell energy and 
momentum dependence of the partial width is then treated using an effective 
Lagrangian model with the required coupling strength adjusted to yield the 
empirical on-shell branching ratio for decay into that channel. Two types of 
decays are considered -- two-body decays in which both decay products are 
stable under the strong interaction, and decays in which one of the decay 
products is itself unstable, so that ultimately more than two decay products 
result.

Decays of the first type all involve the decay of a nucleon resonance into a 
pseudoscalar meson and a spin $\frac{1}{2}$ ground state baryon.  In the 
resonance rest frame, the corresponding widths are given by the expressions
\begin{equation}
\Gamma(\frac{1}{2}^+ \rightarrow \frac{1}{2}^+ + 0^-) = \frac{f^2}{4 \pi}
          \frac{p}{\sqrt{s}}[E_B-m_B],  \label{gam12p}
\end{equation}          
\begin{equation}
\Gamma(\frac{1}{2}^- \rightarrow \frac{1}{2}^+ + 0^-) = \frac{f^2}{4 \pi}
          \frac{p}{\sqrt{s}}[E_B+m_B],  \label{gam12m}
\end{equation}          
\begin{equation}
\Gamma(\frac{3}{2}^+ \rightarrow \frac{1}{2}^+ + 0^-) = \frac{f^2}{12 \pi}
     \frac{p^2}{m_{\pi}^2} \frac{p}{\sqrt{s}}[E_B+m_B],  \label{gam32p}
\end{equation}
\begin{equation}
\Gamma(\frac{3}{2}^- \rightarrow \frac{1}{2}^+ + 0^-) = \frac{f^2}{12 \pi}
     \frac{p^2}{m_{\pi}^2} \frac{p}{\sqrt{s}}[E_B-m_B],  \label{gam32m}
\end{equation}
\begin{equation}
\Gamma(\frac{5}{2}^+ \rightarrow \frac{1}{2}^+ + 0^-) = \frac{f^2}{30 \pi}
     \frac{p^4}{m_{\pi}^4} \frac{p^3}{\sqrt{s}(E_B+m_B)},  \label{gam52p}
\end{equation}
and
\begin{equation}
\Gamma(\frac{5}{2}^- \rightarrow \frac{1}{2}^+ + 0^-) = \frac{f^2}{30 \pi}
     \frac{p^4}{m_{\pi}^4} \frac{p(E_B+m_B)}{\sqrt{s}}  \label{gam52m}
\end{equation}
for the pseudoscalar meson decays, where $p$ is the channel momentum, and 
$m_B$ and $E_B$ are the mass and energy of the baryon decay product.  To obtain
the total contribution to the width from two-body channels at a particular 
energy, the partial widths are summed over all two-body decay channels open at 
that particular energy.

Any part of the full on-shell decay width not accounted for by the two-body 
decays discussed above is attributed to decays in which one of the decay 
products is itself unstable under the strong interaction.  These latter decays 
are approximated as decays into either a ground state baryon and a meson 
resonance or as a decay into a ground state meson and a baryon resonance.  In 
practice, for the low lying nucleon resonances, only decays into the $N\sigma$,
the $N\rho$, and the $\Delta(1232)\pi$ channels are considered.  The $N\sigma$ 
channel is treated as a decay into a nucleon and a scalar meson of zero width.
The corresponding decay widths are the same as those for two-body pseudoscalar 
decays of resonances with the opposite parity.

The two remaining channels both involve a stable ground state hadron and an 
unstable resonance which itself has a width.  To treat such decays, we employ 
the method developed in Ref.~\cite{max1}, which involves an integration over 
the unstable decay product mass of the decay phase space factor multiplied by 
a Breit-Wigner distribution function.  In particular, for decay into either 
channel, the partial width is given by the general expression
\begin{equation}
\Gamma(s) = \frac{g^2}{4\pi} \int_{m_{min}}^{m_{max}} {\cal P}(s,x)
                       {\cal S}(x) dx,   \label{dbut}
\end{equation}
where $g$ is the coupling strength, ${\cal P}$ is the decay phase space factor,
and the integration limits are defined by
\begin{eqnarray}
m_{min} &=& \sqrt{s_{thr}}-m_{stable} \nonumber \\
m_{max} &=& \sqrt{s}-m_{stable}.   \label{limits}
\end{eqnarray}
In the last expressions, $\sqrt{s_{thr}}$ is the threshold value of the 
center-of-momentum energy for decay into that channel, and $m_{stable}$ is the 
stable decay product mass (either $m_N$ or $m_{\pi}$).  The Breit-Wigner 
distribution function has the form
\begin{equation}
{\cal S}(x) = \frac{A}{2\pi}
    \frac{\Gamma_{pr}}{(x-m_C)^2+\frac{1}{4}\Gamma_{pr}^2}    \label{scrs}
\end{equation}
where $\Gamma_{pr}$ is the unstable decay product width, $m_C$ is the mass of 
the unstable decay product at the center of its mass distribution, and the 
parameter $A$ is defined by the normalization requirement
\begin{equation}
\int_{m_{min}}^{\infty} {\cal S}(x) dx = 1.   \label{norm}
\end{equation}

The vertices for decays of spin $\frac{1}{2}$ resonances into the 
$\Delta(1232)\pi$ channel are related to those for decays of spin 
$\frac{3}{2}$ resonances into the $N\pi$ channel by just the interchange of the
initial and final baryon states.  The vertices for decays of spin 
$\frac{3}{2}$ and spin $\frac{5}{2}$ resonances into the   
$\Delta(1232)\pi$ channel each involve two independent couplings, only one of 
which can be fixed by the on-shell partial width.  To avoid this difficulty, 
we keep, in each case, only the coupling of lowest order in the channel 
momentum.  With that proviso, the phase space factors for decays into the 
$\Delta(1232)\pi$ channel are given by 
\begin{equation}
{\cal P}(\frac{1}{2}^+ \rightarrow \frac{3}{2}^+ + 0^-) = \frac{2}{3}
   \frac{p^2}{m_{\pi}^2} \frac{s}{x^2} \frac{p}{\sqrt{s}}(E+x),  
                      \label{gamd12p}
\end{equation}          
\begin{equation}
{\cal P}(\frac{1}{2}^- \rightarrow \frac{3}{2}^+ + 0^-) = \frac{2}{3}
   \frac{p^2}{m_{\pi}^2} \frac{s}{x^2} \frac{p}{\sqrt{s}}(E-x),
                       \label{gamd12m}
\end{equation}          
\begin{equation}
{\cal P}(\frac{3}{2}^+ \rightarrow \frac{3}{2}^+ + 0^-) =  \frac{5}{9} 
    \frac{p}{\sqrt{s}}(E-x),   \label{gamd32p}
\end{equation}
\begin{equation}
{\cal P}(\frac{3}{2}^- \rightarrow \frac{3}{2}^+ + 0^-) = 
    \frac{p}{\sqrt{s}}(E+x),   \label{gamd32m}
\end{equation}
\begin{equation}
{\cal P}(\frac{5}{2}^+ \rightarrow \frac{3}{2}^+ + 0^-) = \frac{1}{3}
    \frac{p^2}{m_{\pi}^2}\frac{p(E+x)}{\sqrt{s}},   \label{gamd52p}
\end{equation}
and
\begin{equation}
{\cal P}(\frac{5}{2}^- \rightarrow \frac{3}{2}^+ + 0^-) = \frac{7}{45}
    \frac{p^2}{m_{\pi}^2}\frac{p^3}{\sqrt{s}(E+x)}  \label{gamd52m}
\end{equation}
where $x$ is the mass that is integrated over in the mass distribution, $p$ is
the channel momentum for that value of $x$, and $E=\sqrt{x^2+p^2}$.
 
Like the decays of spin $\frac{3}{2}$ and spin $\frac{5}{2}$ resonances 
into the $\Delta(1232)\pi$ channel, decays into the $N\rho$ channel generally 
involve vertices with two independent couplings.  For the spin $\frac{1}{2}$ 
resonances decays, we adopt the same procedure as used in Ref.~\cite{max1}.
For the spin $\frac{3}{2}$ and spin $\frac{5}{2}$ resonance decays, we simply 
drop the couplings of higher order in the momenta and energies.  This yields
the phase space factors
\begin{equation}
{\cal P}(\frac{1}{2}^+ \rightarrow \frac{1}{2}^+ + 1^-) =
   \frac{p^2}{x^2} \frac{p}{\sqrt{s}}\frac{(E_+ -E)^2+2x^2}{E_+},  
                                \label{gamr12p}
\end{equation}          
\begin{equation}
{\cal P}(\frac{1}{2}^- \rightarrow \frac{1}{2}^+ + 1^-) =
   \frac{p^2}{x^2} \frac{p}{\sqrt{s}}\frac{(E_- -E)^2+2x^2}{E_-},   
                                \label{gamr12m}
\end{equation}          
\begin{equation}
{\cal P}(\frac{3}{2}^+ \rightarrow \frac{1}{2}^+ + 1^-) = \frac{1}{24}   
    \frac{p^2}{x^2}\frac{p}{\sqrt{s}}\frac{2(E^2+p^2)^2+x^2(E_+ -E)^2
    +3x^2(E_+ +E)^2}{E_+ m_B^2},   \label{gamr32p}
\end{equation}
\begin{equation}
{\cal P}(\frac{3}{2}^- \rightarrow \frac{1}{2}^+ + 1^-) = \frac{1}{24}
     \frac{p^2}{x^2}\frac{p}{\sqrt{s}}\frac{2(E^2+p^2)^2+x^2(E_- -E)^2
    +3x^2(E_- +E)^2}{E_- m_B^2},   \label{gamr32m}
\end{equation}
\begin{equation}
{\cal P}(\frac{5}{2}^+ \rightarrow \frac{1}{2}^+ + 1^-) = \frac{1}{30}
    \frac{p^2}{m_{\pi}^2}\frac{p}{\sqrt{s}}\frac{E_+}{m_B^2}
    (A_+^2+2B_+^2+C_+^2),         \label{gamr52p}
\end{equation}
and
\begin{equation}
{\cal P}(\frac{5}{2}^- \rightarrow \frac{1}{2}^+ + 1^-) =  \frac{1}{30}
    \frac{p^2}{m_{\pi}^2}\frac{p}{\sqrt{s}}\frac{E_+}{m_B^2}
    (A_-^2+2B_-^2+C_-^2),          \label{gamr52m}
\end{equation}
where
\begin{equation}
A_+ = \sqrt{s}-m_B     \nonumber \\
B_+ = \frac{p^2}{E_+}-\frac{A_+}{2}        \nonumber \\
C_+ = \frac{EA_+}{x}+\frac{p^2(E_+ -E)}{xE_+}        \label{abcp}
\end{equation}
and
\begin{equation}
A_- = \frac{p}{E_+}(\sqrt{s}+m_B)     \nonumber \\
B_- = p(1-\frac{\sqrt{s}+m_B}{2})        \nonumber \\
C_- = \frac{p}{x}\frac{E^2+p^2}{E_+}        \label{abcm}
\end{equation}
with $E_+=E_B+m_B$ and $E_-=E_B-m_B$.  

The dynamic width model described above was employed for all of the three 
and four star status $s$-channel resonances used in the fits. For these 
well-established resonances there are generally enough branching ratio data 
that reasonably good estimates for the partial widths on the resonance mass 
shells can be generated.  For the higher energy, less well-established 
resonances, this is generally not the case.  Hence, for these resonances, we 
employ energy-independent widths which are treated as parameters to be varied
in the fits.

\subsection{Evaluation of the matrix elements}

The matrix elements for the reaction $\gamma p\rightarrow K^+\Lambda$ have the 
general structure
\begin{equation}
\bar{u}_{M_{\Lambda}}(p_{\Lambda})\hat{T}u_{M_p}(p_p) = 
       \bar{u}_{M_{\Lambda}}(p_{\Lambda})[\hat{A}+\hat{B}\gamma_5 
       +\hat{C}\gamma^0+\hat{D}\gamma^0\gamma_5]u_{M_p}(p_p),  
                  \label{dme}
\end{equation}
where $p_p$ and $M_p$ are the 4-momentum and spin projection of the proton, 
and $p_{\Lambda}$ and $M_{\Lambda}$ the 4-momentum and spin projection of the 
$\Lambda$.  The operators $\hat{A}$, $\hat{B}$, $\hat{C}$, and $\hat{D}$ 
depend upon the spin and parities associated with the particular contributions 
considered.  Detailed expressions for these operators are given in the 
appendix.

Eq.~(\ref{dme}) can either be evaluated directly or converted to the equivalent
Pauli form,
\begin{equation}
\bar{u}_{M_{\Lambda}}(p_{\Lambda})\hat{T}u_{M_p}(p_p) = 
    N_{\Lambda} N_p \chi_{M_{\Lambda}}^{\dag}[(\hat{A}+\hat{C})
     +(\hat{B}+\hat{D})\sigma\cdot\hat{p}_p 
        + \sigma\cdot\hat{p}_{\Lambda}(\hat{D}-\hat{B}) 
       +\sigma\cdot\hat{p}_{\Lambda}(\hat{C}-\hat{A})\sigma\cdot\hat{p}_p]
          \chi_{M_p}   \label{pme}
\end{equation}
where
\begin{equation}
N = \sqrt{\frac{E+m}{2m}}              \label{dnorm}
\end{equation}
and
\begin{equation}
\hat{p} = \frac{{\bf p}}{E+m}.  \label{momrat}
\end{equation}
The Pauli matrix elements can be evaluated analytically but the procedure
is rather tedious.  Instead, we have evaluated Eq.~(\ref{pme}) numerically. 
As a check on the procedure, an independent code was written to evaluate the 
Dirac matrix elements numerically without recourse to the Pauli reduction given
by Eq.~(\ref{pme}) and the results compared with those of the numerical 
evaluation of Eq.~(\ref{pme}).

\section{details of the fitting procedure}
\label{sec-fitting}

Table~\ref{tab-resonances} lists the $s$ and $u$-channel resonances with three 
or four star status in the particle data tables \cite{pdt} that are included 
in the present fits.
\begin{table}[htbp]
\caption{Well-established resonances considered in the model}
\label{tab-resonances}
\begin{center}
\begin{ruledtabular}
\begin{tabular}{ccc} 
Resonance & $I$ & $J^P$ \\  \hline
$N(1440)$ & $\frac{1}{2}$ & $\frac{1}{2}^+$ \\
$N(1520)$ & $\frac{1}{2}$ & $\frac{3}{2}^-$ \\
$N(1535)$ & $\frac{1}{2}$ & $\frac{1}{2}^-$ \\
$N(1650)$ & $\frac{1}{2}$ & $\frac{1}{2}^-$ \\
$N(1675)$ & $\frac{1}{2}$ & $\frac{5}{2}^-$ \\
$N(1680)$ & $\frac{1}{2}$ & $\frac{5}{2}^+$ \\
$N(1700)$ & $\frac{1}{2}$ & $\frac{3}{2}^-$ \\
$N(1710)$ & $\frac{1}{2}$ & $\frac{1}{2}^+$ \\
$N(1720)$ & $\frac{1}{2}$ & $\frac{3}{2}^+$ \\
$\Lambda(1405)$ & 0 & $\frac{1}{2}^-$ \\
$\Lambda(1520)$ & 0 & $\frac{3}{2}^-$ \\
$\Lambda(1600)$ & 0 & $\frac{1}{2}^+$ \\
$\Lambda(1670)$ & 0 & $\frac{1}{2}^-$ \\
$\Lambda(1690)$ & 0 & $\frac{3}{2}^-$ \\
$\Lambda(1810)$ & 0 & $\frac{1}{2}^+$ \\
$\Lambda(1820)$ & 0 & $\frac{5}{2}^+$ \\
$\Lambda(1830)$ & 0 & $\frac{5}{2}^-$ \\
$\Lambda(1890)$ & 0 & $\frac{3}{2}^+$ \\
$\Lambda(2110)$ & 0 & $\frac{5}{2}^+$ \\
$\Sigma(1385)$ & 1 & $\frac{3}{2}^+$ \\
$\Sigma(1660)$ & 1 & $\frac{1}{2}^+$ \\
$\Sigma(1670)$ & 1 & $\frac{3}{2}^-$ \\
$\Sigma(1750)$ & 1 & $\frac{1}{2}^-$ \\ 
$\Sigma(1775)$ & 1 & $\frac{5}{2}^-$ \\ 
$\Sigma(1915)$ & 1 & $\frac{5}{2}^+$ \\ 
$\Sigma(1940)$ & 1 & $\frac{3}{2}^-$ \\ 
\end{tabular}
\end{ruledtabular}
\end{center}
\end{table}

On-shell branching ratios for the $s$-channel (nucleon) resonances included in
Table~\ref{tab-resonances} are given in Table~\ref{tab-br}.  Where data
exist, the values appearing in the table are the averages of the values 
given in the most recent particle data tables \cite{pdt}.  It should be noted 
that these values differ somewhat from those used in Ref.~\cite{max3} since in 
that earlier reference, data from earlier particle data tables were employed
which differ somewhat from the data in the most recent tables.
After summing the branching ratios obtained from the particle data tables, any 
remaining decay width still not accounted for was assigned to whatever other 
channels are open for that resonance.  For the two-body decay channels, these 
assignments were guided in part by SU(3) symmetry relations.  Previous work by
one of the present authors indicates that the numerical results are not 
strongly sensitive to the details of the dynamical width model employed, 
provided that the total widths are normalized to the empirical width on-shell.
The reader should consult Ref.~\cite{max2} for details.

\begin{table}[htbp]
\caption{On-shell $N^{\star}$ branching ratios}
\label{tab-br}
\begin{ruledtabular}
\begin{tabular}{ccccccc}
Resonance & \multicolumn{3}{c}{Two body channels} 
          & \multicolumn{3}{c}{Three body channels} \\
   & $N\pi$ & $N\eta$ & $\Lambda K$ & $N\sigma$ & $\Delta(1232)\pi$ & $N\rho$
            \\  \hline
$N(1440)$ & 0.65 & & & 0.075 & 0.25 & 0.025 \\
$N(1520)$ & 0.60 & & & & 0.20 & 0.20 \\
$N(1535)$ & 0.44 & 0.515 & & 0.02 & & 0.025 \\
$N(1650)$ & 0.77 & 0.06 & 0.07 & & 0.03 & 0.07 \\
$N(1675)$ & 0.40 & & & & 0.6 &  \\
$N(1680)$ & 0.60 & & & 0.15 & 0.125 & 0.125 \\
$N(1700)$ & 0.10 & & & & 0.80 & 0.10 \\
$N(1710)$ & 0.15 & 0.06 & 0.14 & 0.25 & 0.26 & 0.14 \\
$N(1720)$ & 0.15 & 0.04 & 0.06 & & & 0.75 \\ 
\end{tabular} 
\end{ruledtabular}
\end{table} 

In addition to the resonances listed in Table~\ref{tab-resonances}, four 
additional nucleon resonances, that have two star status in the particle data 
tables, were included.  These are listed in Table~\ref{tab-res2}.

\begin{table}[htbp]
\caption{Two star nucleon resonances considered in the model}
\label{tab-res2}
\begin{ruledtabular}
\begin{tabular}{cc} 
Resonance & $J^P$  \\  \hline
$N(1900)$ & $\frac{3}{2}^+$  \\
$N(2000)$ & $\frac{5}{2}^+$  \\
$N(2080)$ & $\frac{3}{2}^-$  \\
$N(2200)$ & $\frac{5}{2}^-$  \\
\end{tabular}
\end{ruledtabular}
\end{table}

These higher mass $s$-channel resonances were included to improve the fits to
the data at the higher energy end of the kinematic region considered.  
Aside from the empirical evidence for their existences, as reflected in their 
particle data table listings, they are predicted by quark models \cite{quark}, 
and some of them have been included in another recent analysis of 
photoproduction data \cite{bonn}.  In accord with our philosophy to 
incorporate only resonances for which there is
independent empirical evidence, we have {\em not} included the so-called
``missing'' $\frac{3}{2}^-$ resonance at 1900 MeV in our final results.
Although this resonance has been predicted in quark models \cite{quark} and
has been included in several other analyses of strangeness photoproduction
\cite{mb,ghent,bonn}, there has so far been little evidence for its
existence in any other reactions.  We will comment on this further in
Sec.~\ref{sec-results}.

As discussed previously, width data for the resonances listed in
Table~\ref{tab-res2} are extremely limited or non-existent.  For this reason,
no attempt has been made to extend our dynamical off-shell width model to
these resonances.  Instead, their widths are treated as energy-independent 
parameters to be determined in the fits to the data.  The complete set of 
varied parameters thus includes the coupling strength products in all three 
channels and the total widths of the resonances in Table~\ref{tab-res2}. The 
coupling strength products are defined by the relations
\begin{eqnarray}
F_{N^{\star}} &=& e\kappa_{p N^{\star}}g_{\Lambda K N^{\star}},  \nonumber \\
F_{\Lambda^{\star}} &=& e\kappa_{\Lambda\Lambda^{\star}}g_{\Lambda^{\star}Kp}, 
                               \nonumber \\
F_{\Sigma^{\star}} &=& e\kappa_{\Lambda\Sigma^{\star}}g_{\Sigma^{\star}Kp},
                          \label{res12}
\end{eqnarray}
for the ground state baryons and spin $\frac{1}{2}$ resonances in the $s$ and 
$u$-channels, by 
\begin{eqnarray}
G_{N^{\star}}^1 &=& g_1^{p N^{\star}}g_{\Lambda K N^{\star}},  \nonumber \\
G_{N^{\star}}^2 &=& g_2^{p N^{\star}}g_{\Lambda K N^{\star}},  \nonumber \\
G_{\Lambda^{\star}}^1 &=& g_1^{\Lambda\Lambda^{\star}}g_{\Lambda^{\star}Kp}, 
                               \nonumber \\
G_{\Lambda^{\star}}^2 &=& g_2^{\Lambda\Lambda^{\star}}g_{\Lambda^{\star}Kp}, 
                               \nonumber \\
G_{\Sigma^{\star}}^1 &=& g_1^{\Lambda\Sigma^{\star}}g_{\Sigma^{\star}Kp}, 
                             \nonumber \\
G_{\Sigma^{\star}}^2 &=& g_2^{\Lambda\Sigma^{\star}}g_{\Sigma^{\star}Kp}   
                          \label{res32}
\end{eqnarray}
for the spin $\frac{3}{2}$ and spin $\frac{5}{2}$ resonances in the $s$ and 
$u$-channels, by
\begin{eqnarray}
F_K = -e g_{\Lambda Kp}  \label{tgrnd}
\end{eqnarray}
for the ground state kaon in the $t$-channel, and by 
\begin{eqnarray}
G_{K^{\star}}^V &=& g_{\gamma K K^{\star}}g^V_{\Lambda K^{\star}p}, 
                          \nonumber \\
G_{K^{\star}}^T &=& g_{\gamma K K^{\star}}g^T_{\Lambda K^{\star}p}
                          \label{tchres}
\end{eqnarray}
for the $t$-channel kaon resonances, where $e=0.3029$ is the dimensionless 
electric charge.   Note in Eqs.~(\ref{res12}), that the $N^{\star}$, 
$\Lambda^{\star}$, and $\Sigma^{\star}$ subscripts refer to either
the corresponding ground state baryon or a spin $\frac{1}{2}$ resonance.  
For the proton, we also need the charge coupling product.  This is given by 
\begin{eqnarray}
F_{Cp} &=& e g_{\Lambda Kp}.             \label{chcoup}
\end{eqnarray}

The various Born term coupling products can be related to each other through 
SU(3) symmetry relations, SU(2) isospin coupling coefficients, and the 
well-established values for the Baryon ground state magnetic moments.  In
particular, $F_p$, $F_{\Lambda}$, $F_K$, and $F_{Cp}$ satisfy the simple 
relationships
\begin{eqnarray}
F_p &=& \kappa_p F_{Cp},  \nonumber \\
F_{\Lambda} &=& \kappa_{\Lambda} F_{Cp}, \nonumber \\
F_K &=& -F_{Cp}.   \label{grrel}
\end{eqnarray}
For the magnetic moment factors in these relations, we employ the values 
\cite{pdt} $\kappa_p=-1.79$ and $\kappa_{\Lambda}=-0.729$ (note the 
definitions of the $\kappa$'s as given by \ref{phcoup} and \ref{mut}).

The two strong coupling strengths, $g_{\Lambda Kp}$ and $g_{\Sigma Kp}$, can
each be expressed as a product of an SU(3) isoscalar factor and an SU(2) 
Clebsch-Gordon coefficient,
\begin{eqnarray} 
g_{\Lambda Kp} &=& (00\frac{1}{2}\frac{1}{2}\mid\frac{1}{2}\frac{1}{2})
            f_{\Lambda KN}  \nonumber \\
g_{\Sigma Kp} &=& (10\frac{1}{2}\frac{1}{2}\mid\frac{1}{2}\frac{1}{2})
            f_{\Sigma KN}.  \label{strong}
\end{eqnarray}
The SU(3) isoscalar factors appearing here are related by SU(3) symmetry 
\cite{deswart}. In particular, their ratio can be expressed in terms 
of an SU(3) parameter $\alpha$,
\begin{eqnarray}
\frac{f_{\Sigma KN}}{f_{\Lambda KN}} = 
-\frac{1-2\alpha}{1-\frac{2}{3}\alpha},           \label{su3}
\end{eqnarray}
which, in turn, can be fixed by other couplings within the same SU(3) 
multiplets. Using the empirical values for the couplings of pions to the 
$\Sigma$ and $\Lambda$ baryons \cite{dumb} yields the value $\alpha=0.625$. 
Combining Eqs.~(\ref{strong}) and (\ref{su3}) with the coupling product 
definitions, Eq.~(\ref{res12}), we obtain the ratio
\begin{eqnarray}
\frac{F_{\Sigma}}{F_{\Lambda}} = \frac{1}{\sqrt{3}}
        \frac{1-2\alpha}{1-\frac{2}{3}\alpha}
         \frac{\kappa_{\Lambda}}{\kappa_{\Lambda\Sigma}}.   \label{slratio}
\end{eqnarray}
For the transition magnetic moment parameter, we use the particle data table 
value \cite{pdt}, $\kappa_{\Lambda\Sigma}=1.91$, to get 
$\frac{F_{\Sigma}}{F_{\Lambda}}= 0.647$.  The imposition of these relations 
reduces the number of Born parameters to be varied to just one, which we choose
to be the parameter $F_{Cp}$.

The parameter $F_{Cp}$ is also restricted by SU(3) symmetry relations and by 
other considerations.  A recent study by General and Cotanch \cite{gencot} 
based on a generalized Goldberger-Treiman relation in conjunction with the 
Dashen-Weinstein sum rule, arrived at the pair of constraints,
$0.80\leq\frac{g_{K\Sigma N}}{\sqrt{4\pi}}\leq 2.72$ and
$-3.90\leq\frac{g_{K\Lambda N}}{\sqrt{4\pi}}\leq -1.84$. The second constraint
yields an upper limit for $F_{Cp}$ of -1.98.  Since the photoproduction data,
analyzed within the model presented here, seem to favor a small magnitude for 
$F_{Cp}$, we use this upper limit in the first of our fits, which has $F_{Cp}$ 
fixed. For the second fit, we allowed $F_{Cp}$ to assume whatever value yields 
the best fit to the data consistent with the Born term relations given above. 
Comparison of the two fits enabled us to study the extent to which the 
quality of the fit is affected by the constraint on $F_{Cp}$.  Note 
that the choice $F_{Cp}=-1.98$ yields the value $F_{\Sigma}=0.934$, which is
consistent with General and Cotanch's constraint on the value of
$g_{K\Sigma N}$.

The fitting procedure required many iterations starting with a fit of the
most recent CLAS data \cite{newsig} for the unpolarized differential cross
section, given in the c.m.~by
\begin{equation}
\frac{d\sigma}{d\Omega} = \frac{1}{(2\pi)^2}\frac{m_p m_{\Lambda} p_F}
       {4 E_{\gamma} s} \frac{1}{4}\sum_{spins}\mid 
          \langle F\mid\hat{T}\mid I\rangle\mid^2,     \label{gpcross}
\end{equation}
where $p_F$ is the outgoing 3-momentum in the c.m.~and $s=W^2$ is the squared 
c.m.~energy. The resulting parameter values were then employed as starting 
values to fit the cross section data, the CLAS \cite{classig}, SAPHIR
\cite{saphir}, and GRAAL \cite{graal} data for the hyperon polarization
asymmetry $P$, and the CLAS \cite{newpol} data for the double polarization
observables $C_x$ and $C_z$ .
Here, $P$ is defined by  
\begin{eqnarray}
P = \frac{d\sigma_{\Lambda}^+ - d\sigma_{\Lambda}^-} {d\sigma_{\Lambda}^+ +
d\sigma_{\Lambda}^-}, \label{pasy}
\end{eqnarray}
where the superscripts $^+$ and $^-$ refer to spin projections above and below
the scattering plane, and $C_x$ and $C_z$ are defined by 
\begin{equation}
C_{i\prime} = \frac{d\sigma_{\Lambda}^+ - d\sigma_{\Lambda}^-}
          {d\sigma_{\Lambda}^+ + d\sigma_{\Lambda}^-},    \label{dpol}
\end{equation}
where now the superscripts $^+$ and $^-$ refer to $\Lambda$ spin projections 
along and opposite to the $i=z$ or $i=x$ axes, and the incident 
photon is circularly polarized with positive helicity.

At this juncture we generated a prediction for the photon-beam asymmetry,
defined by 
\begin{equation}
 \label{eq-Sigma}
 \Sigma=\frac{d\sigma_\Lambda^\bot-\sigma_\Lambda^\|}
          {d\sigma_\Lambda^\bot+\sigma_\Lambda^\|},
\end{equation}
where $\bot$ and $\|$ refer to polarization vectors perpendicular and parallel
to the scattering plane respectively.  We obtain good agreement with the 
GRAAL \cite{graal} data if we interpret their definition of the observable 
$\Sigma$ to be the negative of the one defined above.  Our definition of this
parameter is the standard one used in most theoretical analyses. The GRAAL 
definition is given in terms of vertical and horizontal planes that are not 
specified relative to the scattering plane, so the possibility exists that 
there is a sign discrepancy between our definition of $\Sigma$ and that of the 
experimentalists.  The last step of the fitting procedure was to include
both the GRAAL and LEPS \cite{leps} $\Sigma$ data in the fit (with
the sign of Eq.~\ref{eq-Sigma} reversed).

In carrying out the fits, we minimized the $\chi^2$ per degree of freedom 
defined by the relation
\begin{equation}
\frac{\chi^2}{\nu} = \sum \frac{(Y_{calc}-Y_{exp})^2}{\sigma^2},  \label{chi} 
\end{equation}
where the sum is over all of the individual data points, $Y_{calc}$ and
$Y_{exp}$ are the calculated and experimental values of the observable, and
$\sigma^2$ is the squared statistical uncertainty in $Y_{exp}$.  The number
of degrees of freedom is given by $\nu=N_{data}-N_{par}$, where $N_{data}$
is the number of data points and $N_{par}$ is the number of parameters in
the fit.

A code based on a modified Marquardt prescription was employed in the fitting 
procedure.  As well as the parameters themselves, this code generates the 
covariance matrix associated with the fit from which well-defined parameter 
uncertainties can be extracted.

\section{Numerical results and discussion}
\label{sec-results}

The parameters from the fit with all of the Born terms constrained (the first 
fit described in the previous section)  are presented in
Table~\ref{tab-results}.  Here the coupling constant products obtained
for all resonances included in the final fit are listed together with the 
total widths obtained for the higher mass nucleon resonances included. Listed
also are the parameter uncertainties obtained from the covariance matrix of
the fit.  These uncertainties measure the sensitivities of the fit to the
corresponding parameters.  A small relative value for this uncertainty means
that the corresponding parameter is well determined by the fit; by contrast, a
large relative value means that the parameter is poorly determined and is
probably strongly correlated with other parameters in the fit.  The $\chi^2$
per degree of freedom associated with this fit is 1.68.  

We have not listed the parameters obtained in the second fit in which
the parameter $F_{Cp}$ was allowed to vary since they are very similar
to those obtained in the fit in which $F_{Cp}$ was fixed at -1.98. In
fact, with the exception of $F_{Cp}$, the differences in the
parameters obtained in the two fits all lie within the parameter
uncertainties derived from the covariance matrices. As mentioned
previously, the values obtained for the $\chi^2$ per degree of freedom
in the two fits differ by less than $1\%$, and the results obtained
for the observables are nearly indistinguishable.  In the second fit,
the value obtained for $F_{Cp}$ increased to the value $-1.76$ with an
uncertainty of 0.16--nearly within one standard deviation of the upper limit.

\begin{table}[htbp]
\caption{Fit results. In this fit the Born parameter $F_{Cp}$ was fixed at 
the value of -1.98.  The width values are given in MeV.}
\label{tab-results}
\begin{ruledtabular}
\begin{tabular}{ccc}
\multicolumn{3}{c}{Spin $\frac{1}{2}$ resonances}  \\\
$N(1440)$ & $F_{N^{\star}}$ &  4.839$\pm$0.224 \\
$N(1535)$ & $F_{N^{\star}}$ &  0.130$\pm$0.027 \\
$N(1650)$ & $F_{N^{\star}}$ &  0.100$\pm$0.012 \\
$N(1710)$ & $F_{N^{\star}}$ &  0.0008$\pm$0.011 \\
$\Lambda(1405)$ & $F_{\Lambda^{\star}}$ &  3.43 $\pm$5.10 \\ 
$\Lambda(1670)$ & $F_{\Lambda^{\star}}$ & -8.70$\pm$6.51 \\
\multicolumn{3}{c}{Spin $\frac{3}{2}$ resonances}  \\
$N(1520)$ & $G_{N^{\star}}^1$ &  0.370$\pm$0.090 \\
          & $G_{N^{\star}}^2$ & -0.067$\pm$0.128 \\
$N(1700)$ & $G_{N^{\star}}^1$ & -0.453$\pm$0.052 \\
          & $G_{N^{\star}}^2$ & -0.391$\pm$0.067 \\
$N(1720)$ & $G_{N^{\star}}^1$ & -0.105$\pm$0.004 \\
          & $G_{N^{\star}}^2$ & -0.200$\pm$0.013 \\
$N(1900)$ & $G_{N^{\star}}^1$ & -0.051$\pm$0.003 \\
          & $G_{N^{\star}}^2$ & -0.050$\pm$0.008 \\
          & $\Gamma$  & 258.6$\pm$9.8 \\
$N(2080)$ & $G_{N^{\star}}^1$ &  0.006$\pm$0.004 \\
          & $G_{N^{\star}}^2$ &  0.003$\pm$0.003 \\
          & $\Gamma$  & 65.5$\pm$21.4 \\
$\Lambda(1890)$ & $G_{\Lambda^{\star}}^1$ & -4.90$\pm$0.59 \\
                & $G_{\Lambda^{\star}}^2$ &  5.12$\pm$4.72 \\
$\Sigma(1385)$ & $G_{\Sigma^{\star}}^1$ &  1.728$\pm$0.414 \\
               & $G_{\Sigma^{\star}}^2$ & -2.14$\pm$3.09 \\
$\Sigma(1940)$  &  $G_{\Sigma^{\star}}^1$ &  0.128$\pm$0.341 \\
               & $G_{\Sigma^{\star}}^2$ & -1.098$\pm$0.714 \\
\multicolumn{3}{c}{Spin $\frac{5}{2}$ resonances}  \\
$N(1675)$ & $G_{N^{\star}}^1$ &  0.0069$\pm$0.0004 \\
          & $G_{N^{\star}}^2$ &  0.0272$\pm$0.0014 \\
$N(1680)$ & $G_{N^{\star}}^1$ & -0.0104$\pm$0.0040 \\
          & $G_{N^{\star}}^2$ &  0.0196$\pm$0.0057 \\
$N(2000)$ & $G_{N^{\star}}^1$ & -0.0130$\pm$0.0069 \\
          & $G_{N^{\star}}^2$ & -0.0272$\pm$0.0127 \\
          & $\Gamma$  & 1133$\pm$490 \\
$N(2200)$ & $G_{N^{\star}}^1$ & -0.0009$\pm$0.0003 \\
          & $G_{N^{\star}}^2$ & -0.0035$\pm$0.0010 \\
          & $\Gamma$  & 371.4$\pm$91.5 \\
$\Lambda(1820)$ & $G_{\Lambda^{\star}}^1$ &  0.388$\pm$0.170 \\
                & $G_{\Lambda^{\star}}^2$ &  0.017$\pm$1.447 \\
$\Lambda(1830)$ & $G_{\Lambda^{\star}}^1$ & -0.555$\pm$0.075 \\
                & $G_{\Lambda^{\star}}^2$ &  1.151$\pm$0.390 \\
$\Lambda(2110)$ & $G_{\Lambda^{\star}}^1$ &  0.127$\pm$0.120 \\
                & $G_{\Lambda^{\star}}^2$ & -0.181$\pm$0.989 \\
$\Sigma(1775)$ & $G_{\Sigma^{\star}}^1$ &  0.517$\pm$0.072 \\
               & $G_{\Sigma^{\star}}^2$ & -1.083$\pm$0.375 \\
$\Sigma(1915)$ & $G_{\Sigma^{\star}}^1$ & -0.526$\pm$0.287 \\
               & $G_{\Sigma^{\star}}^2$ & -0.047$\pm$2.420 \\
\multicolumn{3}{c}{$t$-channel resonances}  \\
$K(892)$ & $G_{K^{\star}}^V$ &  1.090$\pm$0.137 \\
         & $G_{K^{\star}}^T$ & -2.325$\pm$0.338 \\
$K(1270)$ & $G_{K^{\star}}^V$ & 3.074$\pm$0.329 \\
          & $G_{K^{\star}}^T$ & 3.275$\pm$1.350 \\
\end{tabular}
\end{ruledtabular}
\end{table}

Several observations are in order concerning the contents of
Table~\ref{tab-results}.  First, it will be noted that the number of hyperon
resonances appearing in this table is much smaller than the number appearing
in Table~\ref{tab-resonances}.  We found that attempts to incorporate all of
the hyperon resonances listed in Table~\ref{tab-resonances} resulted in
unacceptably large coupling products for many of these resonances
accompanied by enormous parameter uncertainties.  This clearly indicates that
a model that incorporates all of the possible $u$-channel resonances is too
rich; i.e., the photoproduction reaction is not sensitive to particular
$u$-channel contributions and thus, cannot be used to unambiguously determine
individual $u$-channel coupling products.  The reason for this is obvious once
one notes that the kinematic variable $u$ in the photoproduction reaction is
usually negative.  This makes the denominators of the intermediate baryon
propagators in the $u$-channel always large in magnitude and insensitive to
the baryon mass.  Thus, individual contributions to the reaction amplitude in
the $u$-channel are difficult to distinguish from one another and become
highly correlated since many different combinations of the $u$-channel
coupling products yield the same result in the reaction matrix element.  Two
consequences of this are very large parameter uncertainties and the
possibility of a run away effect in the fitting routine in which small
increases in $\chi^2/\nu$ result from simultaneous huge increases in
correlated couplings.
\clearpage

To avoid this difficulty, we systematically removed those $u$-channel
resonances with the largest parameter uncertainties and then refit the
remaining parameters, accepting the result if the resulting $\chi^2/\nu$ did
not increase by more than a few percent over the value obtained with all 
resonances included.  This procedure led to reduced values of the remaining 
$u$-channel coupling products and greatly reduced parameter uncertainties, 
finally culminating in the values listed in Table~\ref{tab-results}.  
Attempts to further reduce the number of $u$-channel resonances incorporated 
in the model led to larger increases in $\chi^2/\nu$ ($10\%$ or more) than we 
deemed acceptable.  As the procedure was carried out, the values obtained for 
the $s$ and $t$-channel coupling products did not shift significantly, 
indicating that these coupling products are not sensitive to the model 
employed for the $u$-channel.

Even with the reduction in the number of $u$-channel resonances included in 
the fit,  the parameter uncertainties associated with the hyperon resonances 
are still rather large, often larger than the magnitudes of the couplings 
themselves.  This indicates that the $u$-channel couplings are still highly 
correlated and rather poorly determined.  Evidently, the photoproduction 
reaction is of limited value as a means for studying the couplings of hyperon 
resonances.

\begin{figure}[tbp]
\begin{center}
\includegraphics[scale=0.8]{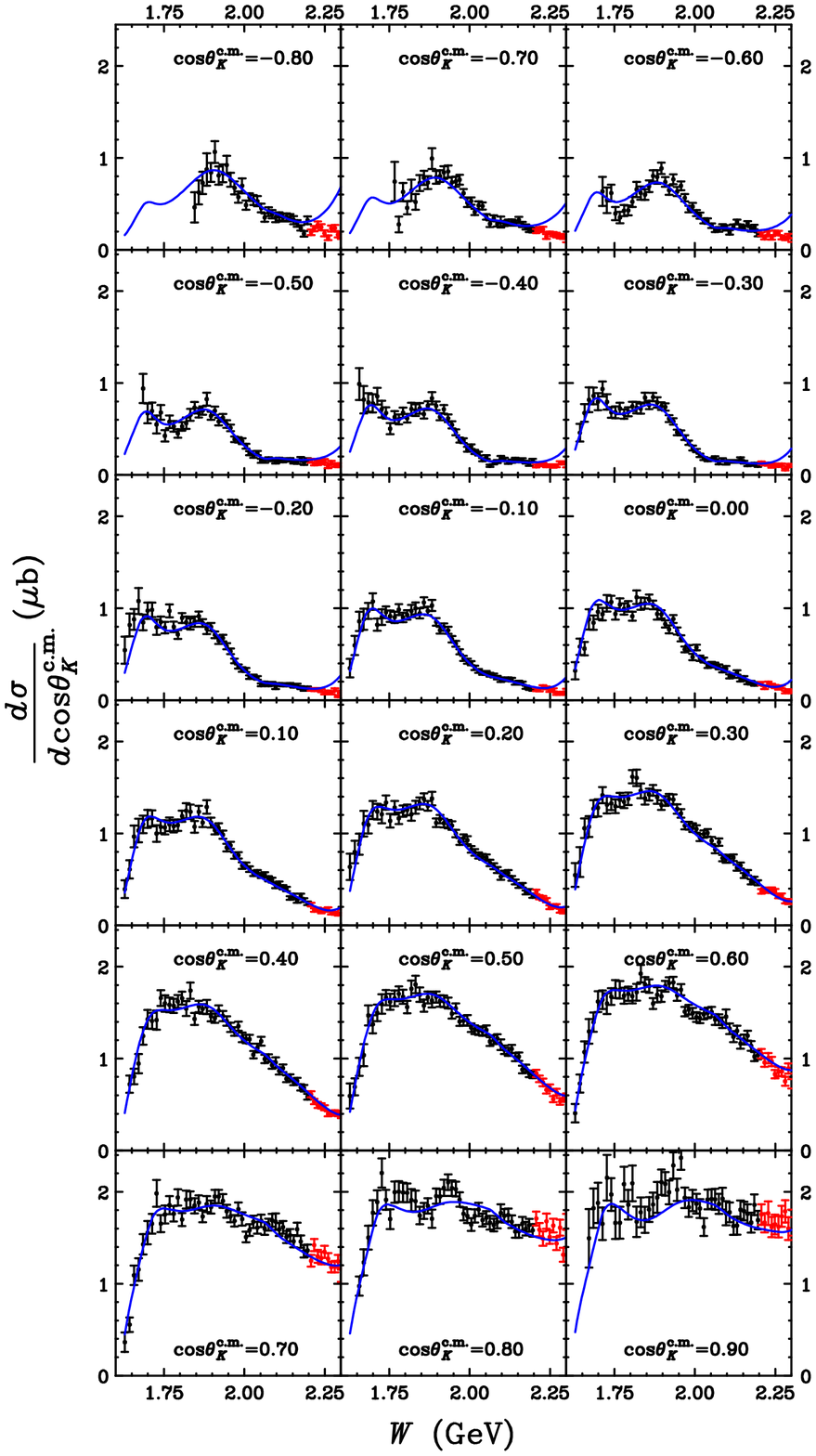}
\caption{(Color online) Differential cross section vs.~$W$ for bins of
$\cos\theta_K^{c.m.}$ as indicated.  Data above 2.2 GeV (red points)
were not included in the fit. The blue curve is from our fit and the data
are from Ref.~\cite{newsig}.}
\label{cscW}
\end{center}
\end{figure}

\begin{figure}[tbp]
\begin{center}
\includegraphics[scale=0.8]{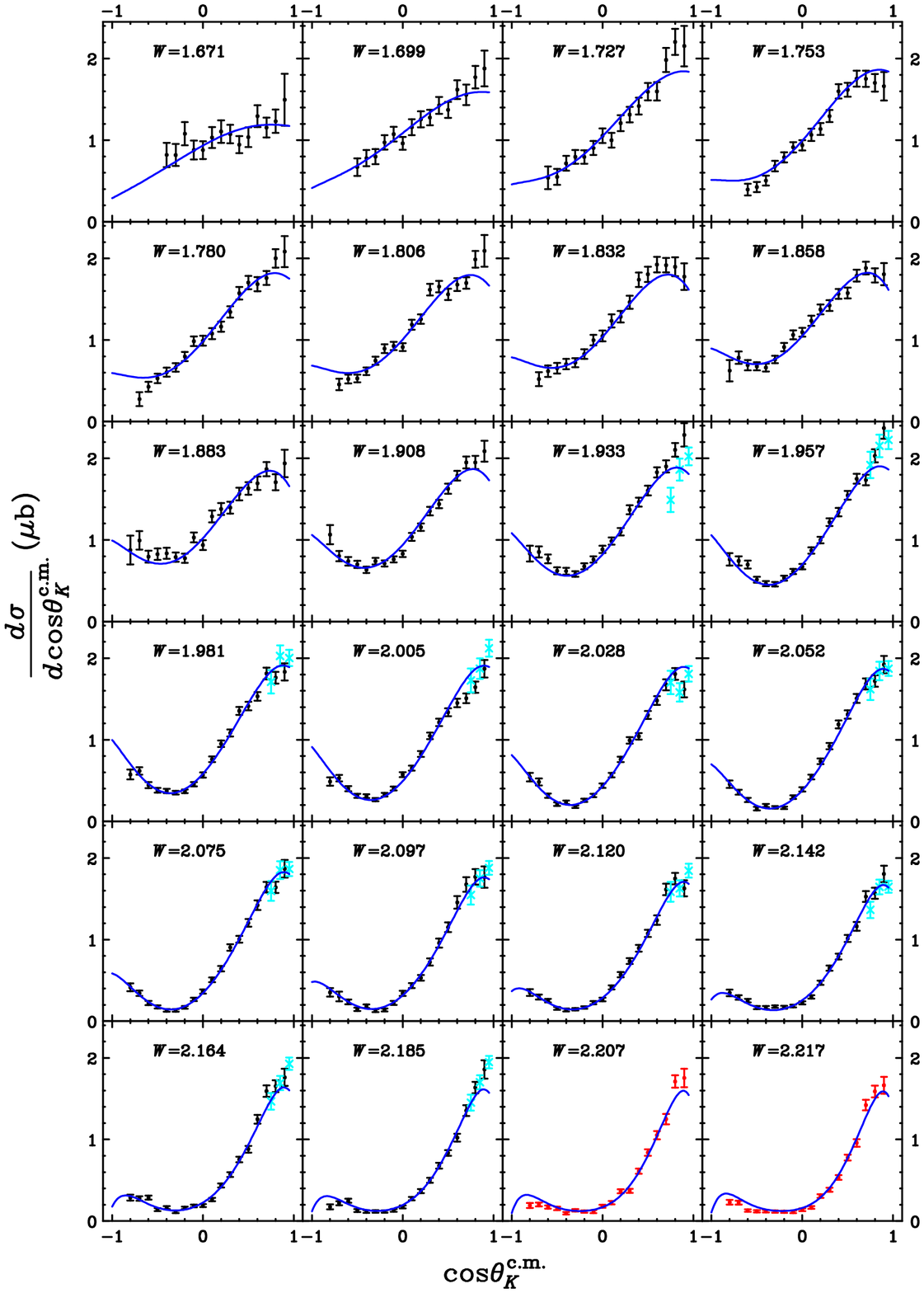}
\caption{(Color online) Differential cross section
vs.~$\cos\theta_K^{c.m.}$ for bins of $W$ as indicated. The blue curve is
from our fit, with black data points from CLAS \cite{newsig} and light blue
points from LEPS \cite{leps}.  The highest two $W$ bins are CLAS data
that were not included in the fit.}
\label{csctheta}
\end{center}
\end{figure}

\begin{figure}[tbp]
\begin{center}
\includegraphics[scale=0.8]{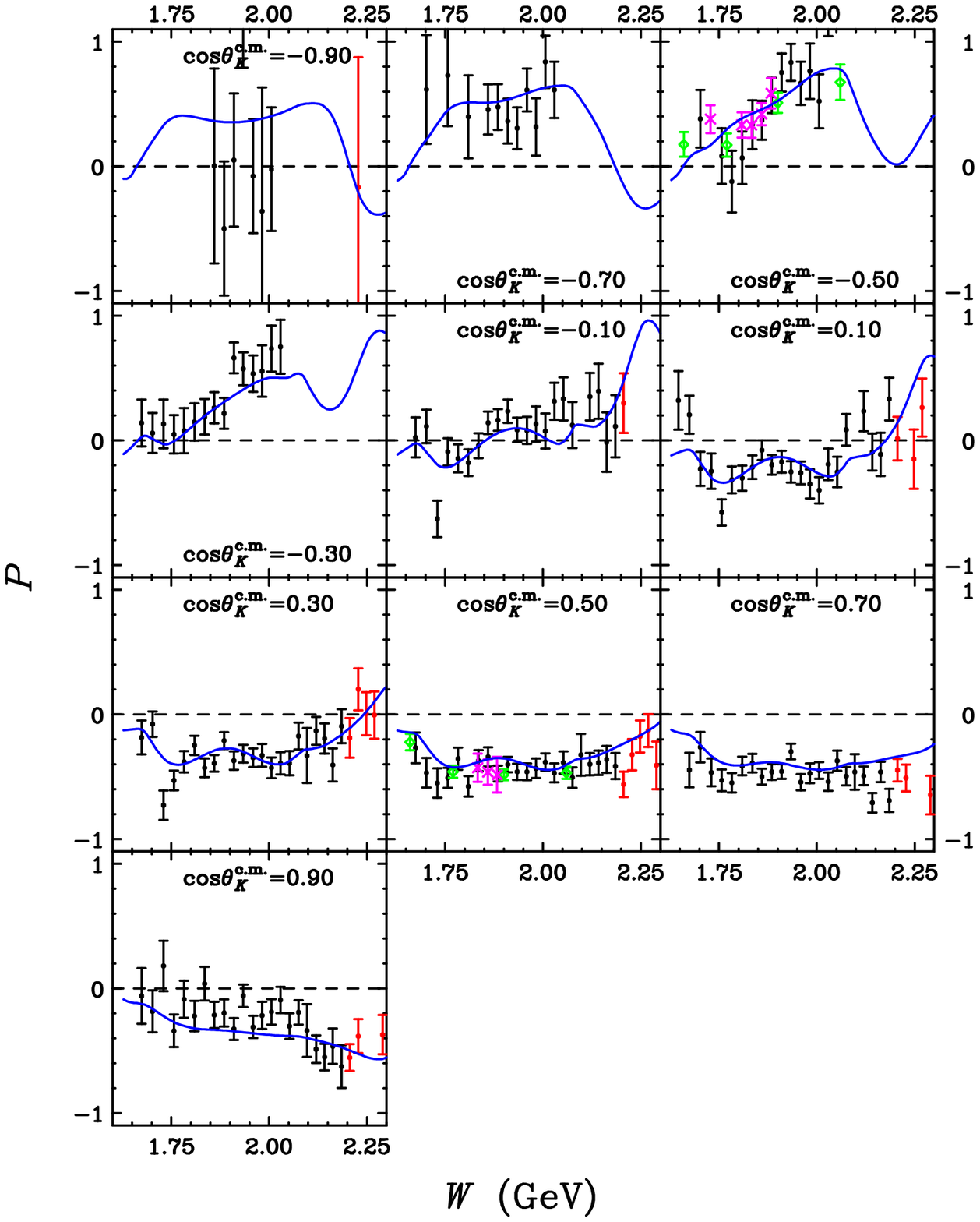}
\caption{(Color online) $\Lambda$ polarization, $P$, vs.~$W$ for bins of
$\cos\theta_K^{c.m.}$ as indicated.  Data above 2.2 GeV (red points)
were not included in the fit. The blue curve is from our fit with black
data points from CLAS \cite{classig}, magenta points from GRAAL
\cite{graal}, and green points from SAPHIR \cite{saphir}.}
\label{PW}
\end{center}
\end{figure}

\begin{figure}[tbp]
\begin{center}
\includegraphics[scale=0.8]{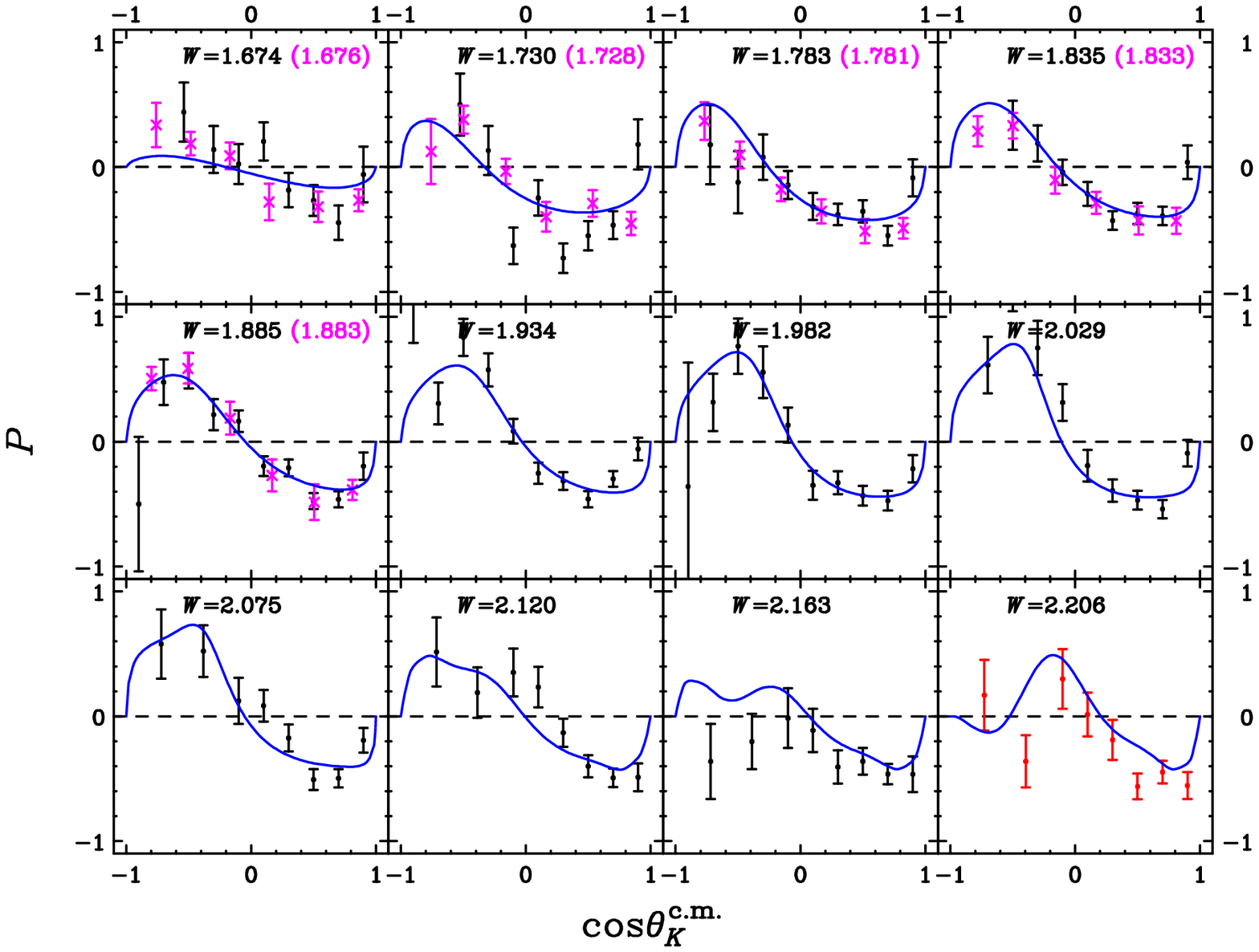}
\caption{(Color online) $\Lambda$ polarization, $P$,  
vs.~$\cos\theta_K^{c.m.}$ for bins of $W$ as indicated.  The blue curve is
from our fit with black data points from CLAS \cite{classig}.  The magenta
points from are from SAPHIR \cite{saphir} with a similar value of $W$,
which is indicated in the parantheses.}
\label{Ptheta}
\end{center}
\end{figure}

\begin{figure}[p]
\begin{center}
\includegraphics[scale=0.8]{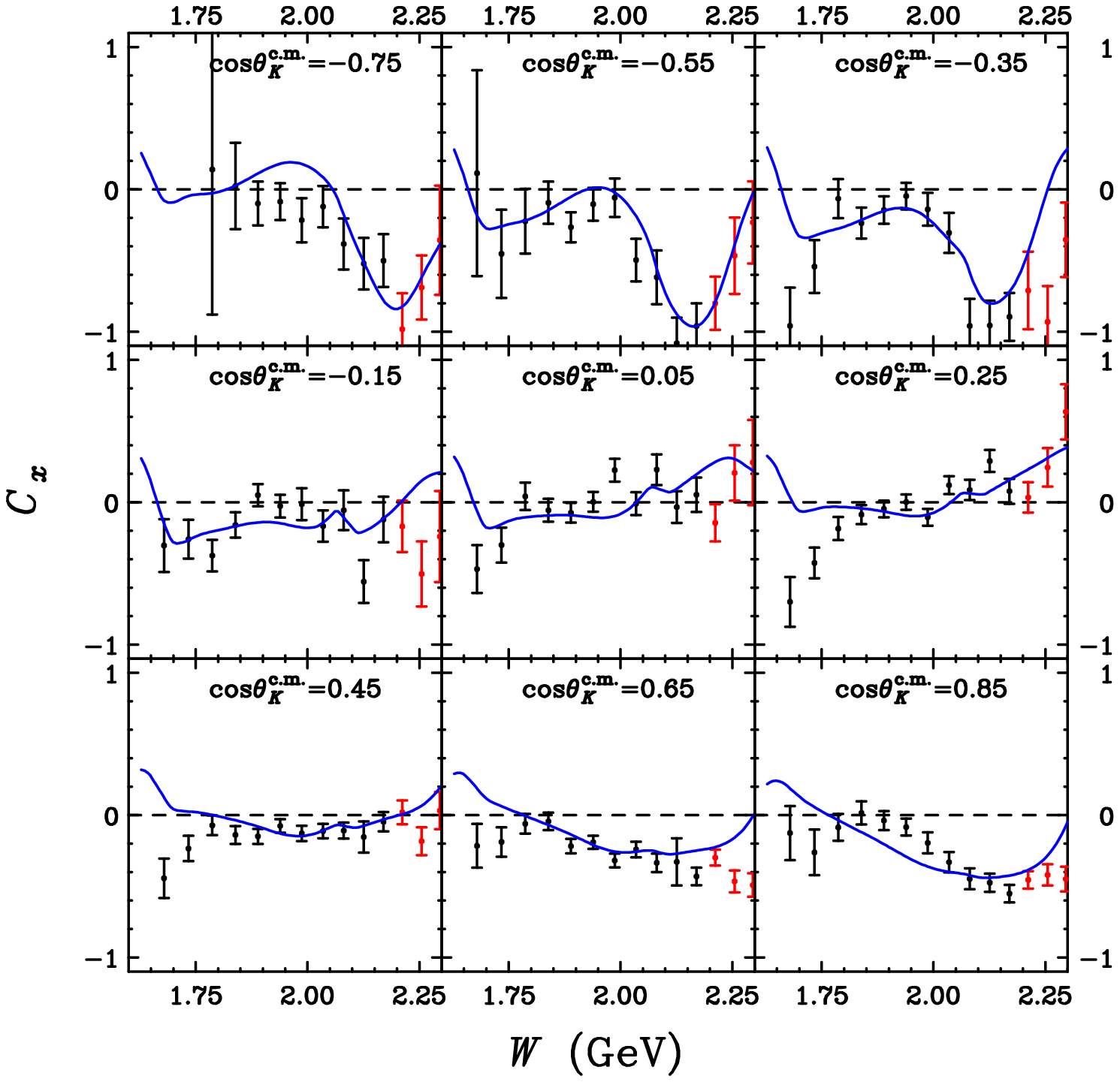}
\caption{(Color online) $C_x$, vs.~$W$ for bins of
$\cos\theta_K^{c.m.}$ as indicated.  Data above 2.2 GeV (red points)
were not included in the fit. The blue curve is from our fit and the data
are from CLAS \cite{newpol}.}
\label{CxW}
\end{center}
\end{figure}

\begin{figure}[tbp]
\begin{center}
\includegraphics[scale=0.8]{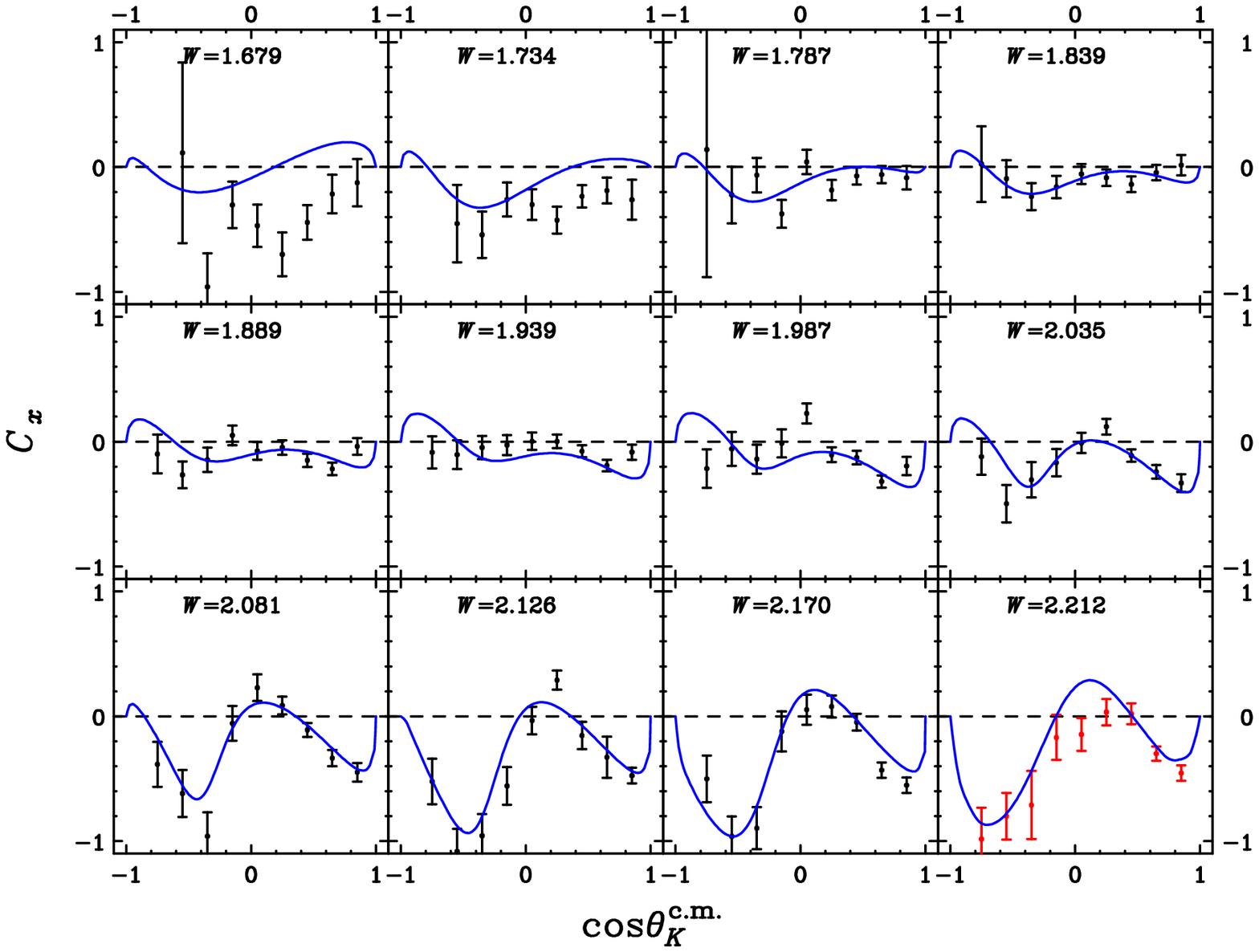}
\caption{(Color online) $C_x$ vs.~$\cos\theta_K^{c.m.}$ for bins of $W$ as
indicated.  The blue curve is from our fit and the data are from
CLAS \cite{newpol}.  The data in the highest $W$ bin were not included
in the fit.}
\label{Cxtheta}
\end{center}
\end{figure}

\begin{figure}[tbp]
\begin{center}
\includegraphics[scale=0.8]{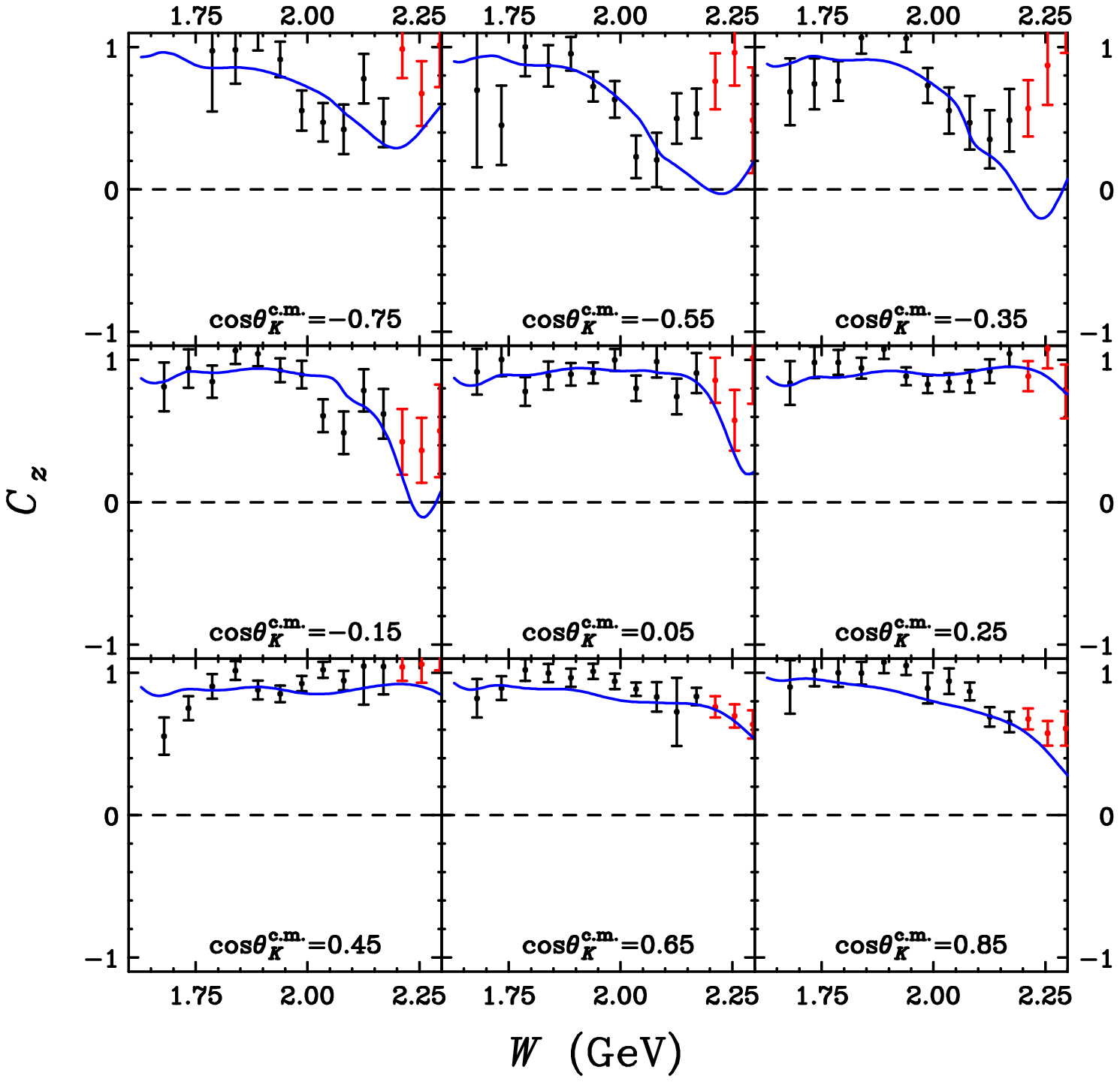}
\caption{(Color online) $C_z$, vs.~$W$ for bins of
$\cos\theta_K^{c.m.}$ as indicated.  Data above 2.2 GeV (red points)
were not included in the fit. The blue curve is from our fit and the data
are from CLAS \cite{newpol}.}
\label{CzW}
\end{center}
\end{figure}

\begin{figure}[tbp]
\begin{center}
\includegraphics[scale=0.8]{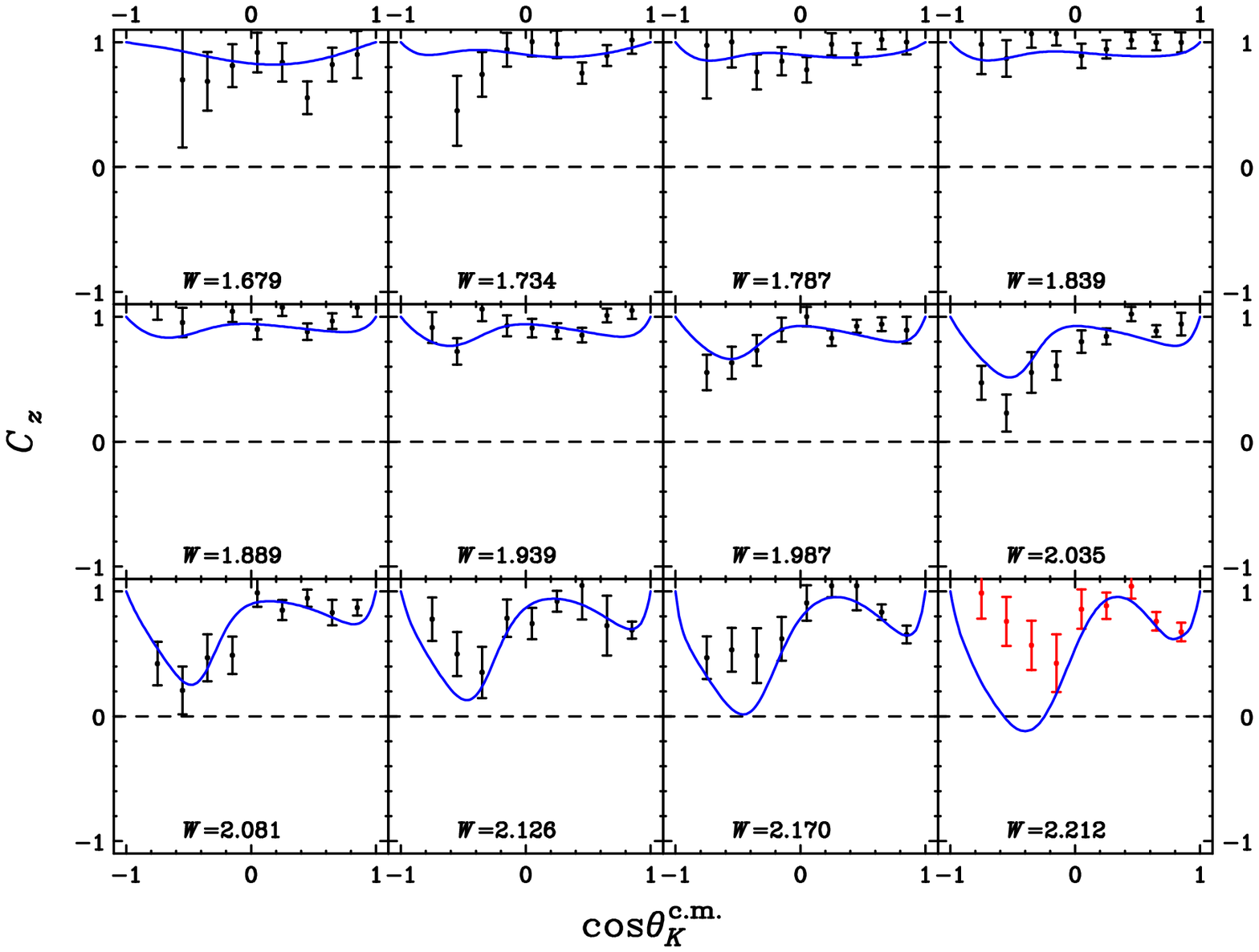}
\caption{(Color online) $C_z$ vs.~$\cos\theta_K^{c.m.}$ for bins of $W$ as
indicated.  The blue curve is from our fit and the data are from CLAS
\cite{newpol}.  The data in the highest $W$ bin were not included
in the fit.} 
\label{Cztheta}
\end{center}
\end{figure}

\begin{figure}[tbp]
\begin{center}
\includegraphics[scale=0.8]{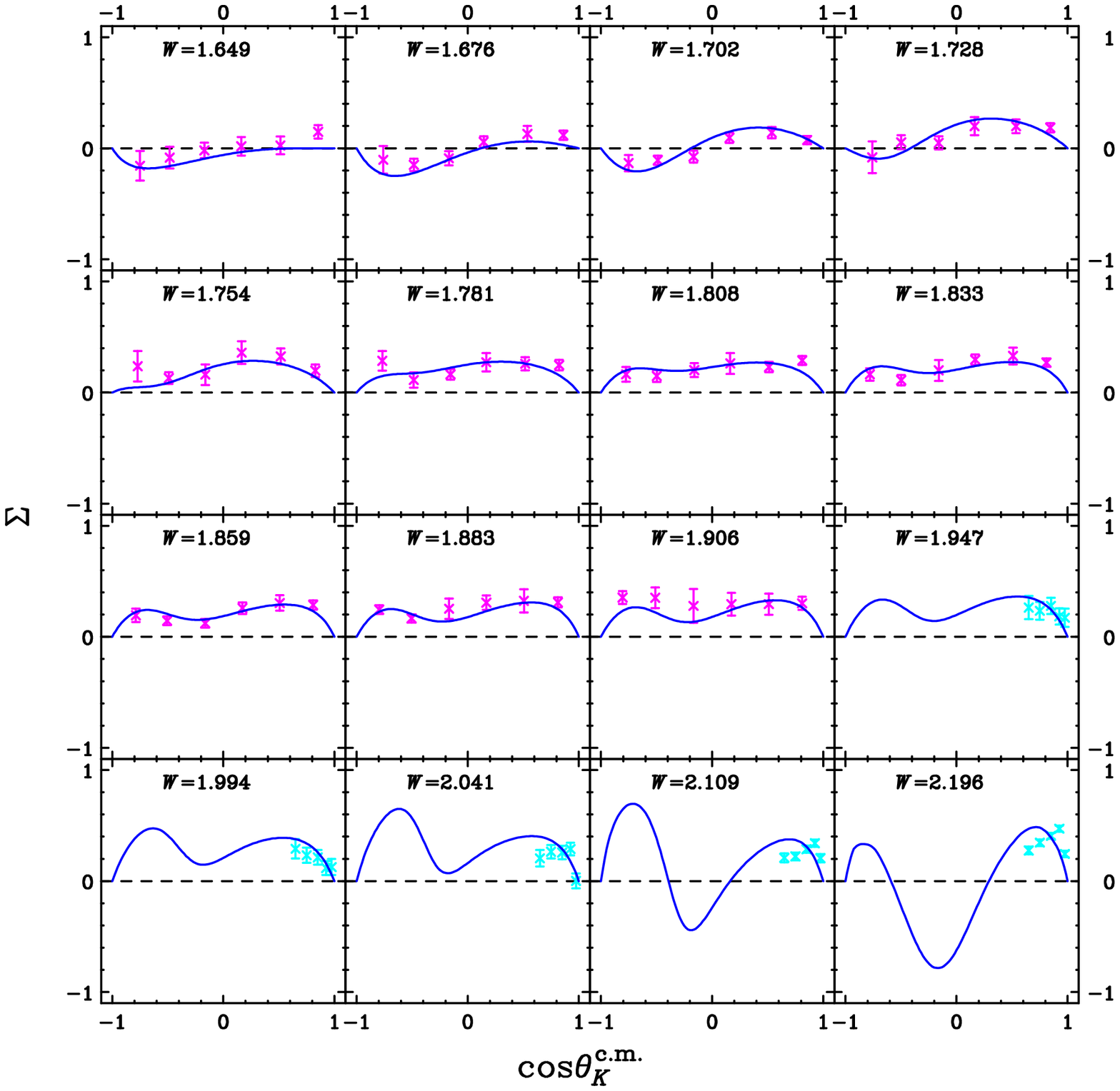}
\caption{(Color online) $\Sigma$ vs.~$\cos\theta_K^{c.m.}$ for bins of $W$ as
indicated.  The blue curve is from our fit and the data are from GRAAL
\cite{graal} (magenta) and LEPS \cite{leps} (light blue).} 
\label{Sigtheta}
\end{center}
\end{figure}

The quality of our fits is illustrated in Figs.~\ref{cscW}-\ref{Sigtheta}.
Fig.~\ref{cscW} shows the $W$ dependence of the differential cross section
for different values of $\cos\theta_K^{c.m.}$. Over the range in $W$ that
the data were fit (up to 2.2 GeV), our fit well reproduces the features of
the data.  The minor exception is in the two forward-angle bins at around
$W\sim 1900$ GeV where the data show a peak not indicated by our fit.  This
could well be an indication of the presence of the $D_{13}(1900)$ included
by others \cite{mb,ghent,bonn}.  Fig.~\ref{csctheta} shows the same cross
section plotted against $\cos\theta_K^{c.m.}$ for a selection of $W$
values.  The figure also includes data from LEPS, which are generally
consistent with the CLAS data.

Our model actually does a good job of matching the cross section data
up to $W\sim 2.3$ GeV for forward angles, a trend seen for the other
observables as well.

The $\Lambda$ polarization fits are shown as a function of $W$ in
Fig.~\ref{PW} and as a function of $\cos\theta_K^{c.m.}$ for selected $W$
bins in Fig.~\ref{Ptheta}. As with the cross section, our fit reproduces the major
features of the data quite well. 

Figs.~\ref{CxW} and \ref{Cxtheta} show the fits of the $C_x$ data.  Over
most of the kinematic range of data, we again reproduce the general
features of the data with our fit.  The exception here is at low $W$ (see
especially Fig.~\ref{Cxtheta}, panel $W=1.679$ GeV), where the data are
systematically lower than our fit.

Figs.~\ref{CzW} and \ref{Cztheta} show the fits of the $C_z$ data.  In the
two most forward angle bins of the $W$ distributions, the fit is
systematically lower than the data from about 1.9 to 2.1 GeV. Otherwise,
the data are well represented by the fit.

Finally, Fig.~\ref{Sigtheta} shows the photon-beam asymmetry $\Sigma$ as a
function of  $\cos\theta_K^{c.m.}$.  The fit shows excellent agreement with
the GRAAL data ($W\leq 1.906$ GeV).  However, from $W=1.947$ GeV up to 2.2
GeV, disagreement between these LEPS data and the fit grows with
$W$.

Detailed examination of the matrix elements in the model reveals that the 
$s$-channel matrix elements, the $u$-channel matrix elements, and the
$t$-channel matrix elements, when taken individually, all increase 
monotonically with energy.  This indicates that successful reproduction of the 
data results from a subtle interference between the matrix elements in 
different channels. Because the Mandelstam variables $u$ and $t$ are 
generally negative over the energy range of interest here, the individual 
resonance contributions in the
$u$ and $t$-channel are only weakly dependent on the energy.  By contrast the 
individual $s$-channel contributions are strongly energy dependent, especially
at energies near the resonance masses. Consequently, to successfully fit the
data within a given energy range with reasonable parameter values, it is
necessary to include nucleon resonances with masses that span the
full energy range considered. Since only nucleon resonances with masses up to
2.2 GeV were included in the fits, attempts to fit the data beyond
2.2 GeV resulted in a $\chi^2/\nu$ that increased rapidly
with $W$.  Furthermore, because of the subtle 
interferences among the three channels that are fine tuned by the fit, fits 
obtained within a given range of energies cannot be extended 
significantly beyond that range.  Indeed, one finds that beyond the energy 
range of the fit, the calculated cross sections increase precipitously with 
energy, in marked contrast with the data.

\section{Conclusions}
\label{sec-conclusion}

In summary, we have presented a new fit of the kaon photoproduction data from
the CLAS collaboration using an effective Lagrangian model.  In this fit
the Born terms were all fixed at values imposed by SU(3) symmetry constraints 
and other considerations. A second fit, in which the value of the Born 
parameter $F_{Cp}$ was allowed to vary while requiring the other Born 
parameters to satisfy SU(3) relations among the various Born couplings, did
not differ significantly from the first fit.  In both
fits, data for the unpolarized cross section, the hyperon polarization 
asymmetry, and two double polarization observables from threshold up to 2.2 GeV
were included.

In general, the new fit yields good representations of both the cross
section data and the spin observables.  One exception is the forward-angle
cross section data around 1.9 GeV.  This suggests that our fit is perhaps
missing one or more resonances in this energy range--a subject of future
work.  Further improvements to our model should result from including other
polarization observables.  The CLAS collaboration expects to produce
multiple spin observables in the near future \cite{klein} including
data with a polarized target and a linearly polarized photon beam.  Such data
should provide more constraints and potentially enable inclusion of
resonances that are not necessary in our present fit.

The results of this work can be employed to study the electroproduction of 
kaons from the proton, i.e., the reaction $ep\rightarrow e^{\prime}K^+\Lambda$.
The latter reaction involves a virtual, rather than a real, photon in the 
strangeness production interaction and thus, requires electromagnetic form 
factors at the photon vertices.  If one uses the photoproduction fits described
here to represent the underlying reaction mechanism in the electroproduction 
reaction, then one can use electroproduction data to study the electromagnetic 
form factors associated with the intermediate hadrons in the different reaction
channels.  Of course, the electroproduction reaction involves longitudinal, as
well as transverse, photons, but the corresponding contributions to the 
reaction amplitude are related through the Lorentz structure of the 
interaction vertices, i.e., through the fact that the photon polarization 
vector is a Lorentz 4-vector.  Thus, it should be possible to extract 
important information concerning the electromagnetic form factors of hadronic 
resonances using the fits described here in conjunction with electroproduction 
data.  Work in this direction is currently in progress.
\clearpage 

\appendix*

\section{amplitude operators}

The operators $\hat{A}$, $\hat{B}$, $\hat{C}$, and $\hat{D}$ appearing in 
Eqs.~(\ref{dme}) and (\ref{pme}) depend on the spin and parity of the 
particular intermediate hadron considered.  They can all be expressed
in terms of a set of $\Sigma$ and $\Omega$ operators 
defined by the relations
\begin{eqnarray}
\Sigma(a,b) &=& a_0b_0-\sigma\cdot{\bf a}\sigma\cdot{\bf b}   \nonumber \\
\Omega(a,b) &=& b_0\sigma\cdot{\bf a}-a_0\sigma\cdot{\bf b},    \label{op2}
\end{eqnarray}
\begin{eqnarray}
\Sigma_3(a,b,c) &=& a_0\Sigma(b,c)-\sigma\cdot{\bf a}\Omega(b,c) \nonumber \\
\Omega_3(a,b,c) &=& a_0\Omega(b,c)-\sigma\cdot{\bf a}\Sigma(b,c), \label{op3}
\end{eqnarray}
and
\begin{eqnarray}
\Sigma_4(a,b,c,d) &=& \Sigma(a,b)\Sigma(c,d)+\Omega(a,b)\Omega(c,d) 
              \nonumber \\
\Omega_4(a,b,c,d) &=& \Sigma(a,b)\Omega(c,d)+\Omega(a,b)\Sigma(c,d),
          \label{op4}
\end{eqnarray}
where $a_0$ and ${\bf a}$ are the time and space components of the 4-vector
$a$. In terms of these operators, the operators for intermediate baryons 
with positive parity and spin $\frac{1}{2}$ are  
\begin{eqnarray}
\hat{A}_s^{\frac{1}{2}^+} &=& FD(p)m\Omega(p_{\gamma},\epsilon),  \nonumber \\
\hat{B}_s^{\frac{1}{2}^+} &=& FD(p)m\Sigma(p_{\gamma},\epsilon),  \nonumber \\
\hat{C}_s^{\frac{1}{2}^+} &=& -FD(p)\Omega_3(p,p_{\gamma},\epsilon),  
                                      \nonumber \\
\hat{D}_s^{\frac{1}{2}^+} &=& -FD(p)\Sigma_3(p,p_{\gamma},\epsilon)  
                                       \label{sop12}
\end{eqnarray}
in the $s$-channel and
\begin{eqnarray}
\hat{A}_u^{\frac{1}{2}^+} &=& FD(p)m\Omega(p_{\gamma},\epsilon),  \nonumber \\
\hat{B}_u^{\frac{1}{2}^+} &=& FD(p) m\Sigma(p_{\gamma},\epsilon),  \nonumber \\
\hat{C}_u^{\frac{1}{2}^+} &=& FD(p)\Omega_3(p_{\gamma},\epsilon,p), 
                                        \nonumber \\
\hat{D}_u^{\frac{1}{2}^+} &=& FD(p)\Sigma_3(p_{\gamma},\epsilon,p)  
                                        \label{uop12}
\end{eqnarray}
in the $u$-channel, where $p_{\gamma}$ and $\epsilon$ are the photon 4-momentum 
and polarization, $m$ and $p$ are the mass and 4-momentum of the intermediate 
baryon, and $D$ is the propagator denominator defined by
\begin{equation}
D(p) = (p^2-m^2+i m\Gamma)^{-1}.     \label{bigd}
\end{equation}
The coupling products $F$ are defined by Eqs.~(\ref{res12}). Note that the 
intermediate baryon width $\Gamma$ in Eq.~(\ref{bigd}) is zero in the Born 
terms. For an intermediate proton there are additional contributions to the 
operators from the charge coupling.  These are given by
\begin{eqnarray}
\hat{A}^{charge} &=& eg_{\Lambda Kp}D(p)\Omega(p,\epsilon),  \nonumber \\
\hat{B}^{charge} &=& eg_{\Lambda Kp}D(p)\Sigma(p,\epsilon),  \nonumber \\
\hat{C}^{charge} &=& eg_{\Lambda Kp}D(p)m\sigma\cdot\epsilon,  \nonumber \\
\hat{D}^{charge} &=& 0.             \label{prot12}
\end{eqnarray}

For contributions with intermediate spin $\frac{3}{2}$ resonances, we define
the coupling parameters
\begin{eqnarray}
\beta_1 &=& F_1+F_2,   \nonumber  \\
\beta_2 &=& F_2-2F_1,  \nonumber  \\
\beta_3 &=& 3F_1-F_2    \label{beta12}
\end{eqnarray}
with
\begin{eqnarray}
F_1 &=& \frac{G^1}{2m_Bm_{\pi}}D(p),  \nonumber  \\
F_2 &=& \frac{mG^2}{(2m_B)^2 m_{\pi}}D(p),  \label{coup32}
\end{eqnarray}
where $m_B$ is the mass of the ground state baryon at the photon vertex, and
$G^1$ and $G^2$ are the couplings defined by Eqs.~(\ref{res32}).  With these 
definitions, the operators for intermediate resonances of positive parity and
spin $\frac{3}{2}$ are given by
\begin{eqnarray}
\hat{A}_s^{\frac{3}{2}^+} &=& \frac{1}{3}[\beta_1\Omega(p_K,k_1)
     +2 F_1(p_K\cdot p)\Omega(p_{\gamma},\epsilon)-3\Omega(p,q_1)
     -2 F_1\Omega_4(p,p_K,p_{\gamma},\epsilon)],  \nonumber \\
\hat{B}_s^{\frac{3}{2}^+} &=& \frac{1}{3}[\beta_1\Sigma(p_K,k_1)
     +2 F_1(p_K\cdot p)\Sigma(p_{\gamma},\epsilon)-3\Sigma(p,q_1)
     -2 F_1\Sigma_4(p,p_K,p_{\gamma},\epsilon)-3 F_2(p_K\cdot k_1)],
      \nonumber \\
\hat{C}_s^{\frac{3}{2}^+} &=& \frac{1}{3m}[\beta_1\Omega_3(p,p_K,k_1)
     +2 F_1(p_K\cdot p)\Omega_3(p,p_{\gamma},\epsilon)
     +3 F_2(p_K\cdot k_1)\sigma\cdot{\bf p}      \nonumber \\
  & & +3m^2\sigma\cdot{\bf q}_1-2m^2 F_1\Omega_3(p_K,p_{\gamma},\epsilon)],  
                      \nonumber \\
\hat{D}_s^{\frac{3}{2}^+} &=& \frac{1}{3m}[\beta_1\Sigma_3(p,p_K,k_1)
     +2 F_1(p_K\cdot p)\Sigma_3(p,p_{\gamma},\epsilon)
     -3 F_2(p_K\cdot k_1)E  \nonumber \\
  & & -3m^2 q_1^0-2m^2 F_1\Sigma_3(p_K,p_{\gamma},\epsilon)]  \label{sop32}
\end{eqnarray}
in the $s$-channel and
\begin{eqnarray}
\hat{A}_u^{\frac{3}{2}^+} &=& \frac{1}{3}[-\beta_3\Omega(k_1,p_K)
     -2 F_1(p_K\cdot p)\Omega(\epsilon,p_{\gamma})+3\Omega(p,q_1)
     +2 F_1\Omega_4(p,\epsilon,p_{\gamma},p_K)],  \nonumber \\
\hat{B}_u^{\frac{3}{2}^+} &=& \frac{1}{3}[-\beta_3\Sigma(k_1,p_K)
     -2 F_1(p_K\cdot p)\Sigma(\epsilon,p_{\gamma})+3\Sigma(p,q_1)
     +2 F_1\Sigma_4(p,\epsilon,p_{\gamma},p_K)-3\beta_2(p_K\cdot k_1)],
       \nonumber \\  
\hat{C}_u^{\frac{3}{2}^+} &=& \frac{1}{3m}[-\beta_1\Omega_3(p,k_1,p_K)
     -2 F_1(p_K\cdot p)\Omega_3(p,\epsilon,p_{\gamma})
     -3 F_2(p_K\cdot k_1)\sigma\cdot{\bf p}     \nonumber \\
  & & -3 m^2\sigma\cdot{\bf q}_2+2m^2 F_1\Omega_3(\epsilon,p_{\gamma},p_K)],  
                                 \nonumber \\
\hat{D}_u^{\frac{3}{2}^+} &=& \frac{1}{3m}[-\beta_1\Sigma_3(p,k_1,p_K)
     -2 F_1(p_K\cdot p)\Sigma_3(p,\epsilon,p_{\gamma})+3 F_2(p_K\cdot k_1)E
                                 \nonumber \\
  & & +3 m^2 q_2^0+2m^2 F_1\Sigma_3(\epsilon,p_{\gamma},p_K)]  \label{uop32}
\end{eqnarray}
in the $u$-channel, where $E$ is the energy of the intermediate resonance, $p_K$
is the kaon 4-momentum,
\begin{eqnarray}
k_1 &=& (p\cdot\epsilon)p_{\gamma}-(p\cdot p_{\gamma})\epsilon, \nonumber  \\  
k_2 &=& (p_K\cdot\epsilon)p_{\gamma}-(p_K\cdot p_{\gamma})\epsilon,  
                        \label{littk}
\end{eqnarray}
and
\begin{eqnarray}
q_1 &=& F_1 k_2 +\beta_2\frac{p_K\cdot p}{3m^2}k_1,  \nonumber  \\
q_2 &=& F_1 k_2-F_2\frac{p_K\cdot p}{3m^2}k_1.           \label{littq}
\end{eqnarray}

For contributions with intermediate spin $\frac{5}{2}$ resonances, we define
the coupling parameters
\begin{eqnarray}
F_1 &=& \frac{G^1}{2m_B(m_{\pi})^3}D(p),  \nonumber  \\
F_2 &=& \frac{mG^2}{(2m_B)^2(m_{\pi})^3}D(p),  \label{coup52}
\end{eqnarray}
where $G^1$ and $G^2$ are the coupling products given by Eqs.~(\ref{res32}), 
and the linear combinations
\begin{eqnarray}
\xi_1 &=& b_1 p_{\gamma}-b_2\epsilon,  \nonumber \\
\xi_2 &=& a_1 p_{\gamma}-a_2\epsilon,  \nonumber \\
\zeta &=& q\cdot\epsilon p_{\gamma}+q\cdot p_{\gamma}\epsilon,  \label{cmom5}
\end{eqnarray}
where
\begin{eqnarray}
a_1 &=& 2q\cdot p_{\gamma}p_B\cdot\epsilon -q\cdot\epsilon p_B\cdot p_{\gamma},
                            \nonumber \\
a_2 &=& q\cdot p_{\gamma}p_B\cdot p_{\gamma},    \nonumber \\
b_1 &=& q\cdot p_{\gamma}p\cdot\epsilon +q\cdot\epsilon p\cdot p_{\gamma},
                            \nonumber \\
b_2 &=& 2q\cdot p_{\gamma}p\cdot p_{\gamma}      \label{ab5}
\end{eqnarray}
with
\begin{equation}
q = p_K-\beta p            \label{q5}
\end{equation}
and
\begin{equation}
\beta = \frac{p\cdot p_K}{m^2}.          \label{beta5}
\end{equation}
Four other useful combinations are
\begin{eqnarray}
c_1 &=& q\cdot\epsilon p_K\cdot p_{\gamma}+q\cdot p_{\gamma}p_K\cdot\epsilon
       -p_K\cdot p_{\gamma}p_K\cdot\epsilon 
       +\frac{1}{5}\beta_K p\cdot\epsilon p\cdot p_{\gamma}, \nonumber \\
c_2 &=& (2q\cdot p_{\gamma}-p_K\cdot p_{\gamma})p_K\cdot p_{\gamma}
       +\frac{1}{5}\beta_K (p\cdot p_{\gamma})^2,   \nonumber \\
c_3 &=& \frac{a_1 p\cdot p_{\gamma}- a_2 p\cdot\epsilon}{m^2}, \nonumber \\
c_4 &=& a_1 p_K\cdot p_{\gamma}-a_2 p_K\cdot\epsilon+p_K\cdot p_{\gamma}
  (p_B\cdot p_{\gamma}p_K\cdot\epsilon-p_B\cdot\epsilon p_K\cdot p_{\gamma})
                               \nonumber \\
 & &   +\frac{1}{5}\beta_K p\cdot p_{\gamma}
   (p\cdot p_{\gamma}p_B\cdot\epsilon -p\cdot\epsilon p_B\cdot p_{\gamma})
                \label{c5}
\end{eqnarray}
with
\begin{equation}
\beta_K = \frac{m_K^2+4(\beta m)^2}{m^2}.  \label{betk}
\end{equation}
In terms of these quantities, we have for positive parity spin $\frac{5}{2}$ 
resonances
\begin{eqnarray}
\hat{A}_s^{\frac{5}{2}^+} &=& F_1[c_1\Omega(p,p_{\gamma})
   -c_2\Omega(p,\epsilon)]+\frac{1}{5}F_2[c_3\Omega(q,p)-\Omega(q,\xi_2)]
                          \nonumber \\
  & &   +\frac{1}{5}F_1[\frac{1}{m^2}\Omega_4(p,q,p,\xi_1)
        +2q\cdot p_{\gamma}\Omega_4(p,q,p_{\gamma},\epsilon)
        -\Omega_4(p,q,\zeta,p_{\gamma})],               \nonumber \\
\hat{B}_s^{\frac{5}{2}^+} &=& F_2c_4+F_1[c_1\Sigma(p,p_{\gamma})
   -c_2\Sigma(p,\epsilon)]+\frac{1}{5}F_2[c_3\Sigma(q,p)-\Sigma(q,\xi_2)]
                          \nonumber \\
  & &   +\frac{1}{5}F_1[\frac{1}{m^2}\Sigma_4(p,q,p,\xi_1)
        +2q\cdot p_{\gamma}\Sigma_4(p,q,p_{\gamma},\epsilon)
        -\Sigma_4(p,q,\zeta,p_{\gamma})],               \nonumber \\ 
\hat{C}_s^{\frac{5}{2}^+} &=& F_2\frac{c_4}{m}\sigma\cdot p
    +F_1m[c_1\sigma\cdot p_{\gamma}-c_2\sigma\cdot\epsilon]-\frac{1}{5m}F_2
    [c_3\Omega_3(p,q,p)-\Omega_3(p,q,\xi_2)] \nonumber \\
  & &   -\frac{1}{5}F_1[\frac{1}{m}\Omega_3(q,p,\xi_1)+2q\cdot p_{\gamma}m
    \Omega_3(q,p_{\gamma},\epsilon)-m\Omega_3(q,\zeta,p_{\gamma})], 
                             \nonumber \\  
\hat{D}_s^{\frac{5}{2}^+} &=& -F_2c_4\frac{E}{m}-F_1c_1mE_{\gamma}
    -\frac{1}{5m}F_2[c_3\Sigma_3(p,q,p)-\Sigma_3(p,q,\xi_2)] 
                \nonumber \\
  & &   -\frac{1}{5}F_1[\frac{1}{m}\Sigma_3(q,p,\xi_1)+2q\cdot p_{\gamma}m
    \Sigma_3(q,p_{\gamma},\epsilon)-m\Sigma_3(q,\zeta,p_{\gamma})]
                                  \label{sop52}
\end{eqnarray}
in the $s$-channel and
\begin{eqnarray}
\hat{A}_u^{\frac{5}{2}^+} &=& F_1[c_1\Omega(p_{\gamma},p)
   -c_2\Omega(\epsilon,p)]+\frac{1}{5}F_2[c_3\Omega(p,q)-\Omega(\xi_2,q)]
                          \nonumber \\
  & &   +\frac{1}{5}F_1[\frac{1}{m^2}\Omega_4(\xi_1,p,p,q)
        +2q\cdot p_{\gamma}\Omega_4(\epsilon,p,p_{\gamma},q)
        -\Omega_4(p_{\gamma},p,\zeta,q)],               \nonumber \\
\hat{B}_u^{\frac{5}{2}^+} &=& F_2c_4+F_1[c_1\Sigma(p_{\gamma},p)
   -c_2\Sigma(\epsilon,p)]+\frac{1}{5}F_2[c_3\Sigma(p,q)-\Sigma(\xi_2,q)]
                          \nonumber \\
  & &   +\frac{1}{5}F_1[\frac{1}{m^2}\Sigma_4(\xi_1,p,p,q)
        +2q\cdot p_{\gamma}\Sigma_4(\epsilon,p,p_{\gamma},q)
        -\Sigma_4(p_{\gamma},p,\zeta,q)],               \nonumber \\
\hat{C}_u^{\frac{5}{2}^+} &=& -F_2\frac{c_4}{m}\sigma\cdot p
    +F_1m[c_2\sigma\cdot\epsilon-c_1\sigma\cdot p_{\gamma}]+\frac{1}{5m}F_2
    [c_3\Omega_3(p,p,q)-\Omega_3(p,\xi_2,q)] \nonumber \\
  & &   +\frac{1}{5}F_1[\frac{1}{m}\Omega_3(\xi_1,p,q)+2q\cdot p_{\gamma}m
    \Omega_3(\epsilon,p_{\gamma},q)-m\Omega_3(p_{\gamma},\zeta,q)], 
                             \nonumber \\  
\hat{D}_u^{\frac{5}{2}^+} &=& F_2c_4\frac{E}{m}+F_1c_1mE_{\gamma}
    +\frac{1}{5m}F_2[c_3\Sigma_3(p,p,q)-\Sigma_3(p,\xi_2,q)] 
                        \nonumber \\
  & &   +\frac{1}{5}F_1[\frac{1}{m}\Sigma_3(\xi_1,p,q)+2q\cdot p_{\gamma}m
    \Sigma_3(\epsilon,p_{\gamma},q)-m\Sigma_3(p_{\gamma},\zeta,q)]
                                  \label{uop52}
\end{eqnarray}
in the $u$-channel.

For intermediate baryons of negative parity, the $\hat{A}$ and $\hat{B}$ 
operators are given by the same expressions as for intermediate baryons of 
positive parity and the same spin; whereas, the $\hat{C}$ and $\hat{D}$
operators are given by expressions that are the negatives of the corresponding
positive parity expressions. 

For the $t$-channel, we define the coupling parameters
\begin{eqnarray}
\alpha^V &=& \frac{G_{K^{\star}}^V}{m_{sc}}D(p),  \nonumber \\
\alpha^T &=& \frac{G_{K^{\star}}^T}{m_{sc}(m_p+m_{\Lambda})}D(p) \label{alfkst}
\end{eqnarray}
where $m_{sc}$ is the same scaling mass that appears in Eqs.~(\ref{tphk*}) and
(\ref{tphk1}),and the $G_{K^{\star}}$ are the coupling products defined by 
Eqs.~(\ref{tchres}). In terms of these parameters, the $t$-channel operators are
given by 
\begin{eqnarray}
\hat{A}_K^t &=& 0,                           \nonumber \\
\hat{B}_K^t &=& eg_{\Lambda K p}D(p)         \nonumber \\
\hat{C}_K^t &=& 0,                           \nonumber \\
\hat{D}_K^t &=& 0                            \label{tkaon}
\end{eqnarray}
for an intermediate kaon,
\begin{eqnarray}
\hat{A}_{K^{\star}}^t &=& i\alpha^T(Ef-\sigma\cdot{\bf p}\sigma\cdot{\bf \xi}),
                          \nonumber \\
\hat{B}_{K^{\star}}^t &=& -i\alpha^T(E\sigma\cdot{\bf \xi}-
                 f\sigma\cdot{\bf p}),   \nonumber \\
\hat{C}_{K^{\star}}^t &=& i\alpha^Vf,    \nonumber \\
\hat{D}_{K^{\star}}^t &=& -i\alpha^V\sigma\cdot{\bf \xi}         \label{tkstar}
\end{eqnarray}
for an intermediate $K^{\star}(892)$ resonance, and 
\begin{eqnarray}
\hat{A}_{K1}^t &=& \alpha^T[\epsilon\cdot p_K \Omega(p,p_{\gamma})
        +E p_{\gamma}\cdot p_K\sigma\cdot{\bf \epsilon}],  \nonumber \\
\hat{B}_{K1}^t &=& \alpha^T[\epsilon\cdot p_K \Sigma(p,p_{\gamma})
        +p_{\gamma}\cdot p_K\sigma\cdot{\bf p}
        \sigma\cdot{\bf \epsilon}],  \nonumber \\
\hat{C}_{K1}^t &=& \alpha^V[p_{\gamma}\cdot p_K\sigma\cdot{\bf \epsilon}
        -\epsilon\cdot p_K\sigma\cdot{\bf p}_{\gamma}],    \nonumber \\
\hat{D}_{K1}^t &=& \alpha^V\epsilon\cdot p_K E_{\gamma}   \label{tk1}
\end{eqnarray}
for an intermediate $K1(1270)$ resonance, where p and E are the 4-momentum
and energy of the intermediate meson, 
\begin{equation}
f = {\bf \epsilon}\cdot{\bf p}_{\gamma}\times{\bf p}_K,        \label{littlef} 
\end{equation}
and
\begin{equation}
{\bf \xi} = {\bf \epsilon}\times(E_K{\bf p}_{\gamma}-E_{\gamma}{\bf p}_K). 
       \label{vecxi} 
\end{equation}


\begin{thebibliography}{99}

\bibitem{thom} H. Thom, Phys. Rev. {\bf 151}, 1322 (1966)
\bibitem{renard} F.M. Renard and Y. Renard, Nucl. Phys. {\bf B1}, 389 (1967);
     F.M. Renard and Y. Renard, Phys. Lett. {\bf 24B}, 159 (1967); F.M. Renard 
     and Y. Renard, Nucl. Phys. {\bf B25}, 490 (1971); Y. Renard, {\em ibid.}
     {\bf B40}, 499 (1972). 
\bibitem{early} H. Thom, E. Gabathuler, D. Jones, B.D. McDaniel, and W.M. 
    Woodward, Phys. Rev. Lett. {\bf 11}, 433 (1963); M. Grilli, L. Mezzetti, 
    M. Nigro, and E. Schiavuta, Nuovo Cim. {\bf 38}, 1467 (1965); D.E. Groom 
    and J.H. Marshall, Phys. Rev. {\bf 159}, 1213 (1967); T. Fujii et al., 
    Phys. Rev. D {\bf 2}, 439 (1970).  
\bibitem{saphir} M.Q. Tran {\em et al.}, Phys. Lett. {\bf B445}, 20 (1998);
     S. Goers {\em et al.}, {\em ibid.} {\bf B464}, 331 (1999); 
     K.H. Glander {\em et al.}, Eur. Phys. J. A {\bf 19}, 251 (2004).
\bibitem{leps} R.G.T. Zegers {\em et al.}, Phys. Rev. Lett. {\bf 91}, 092001
     (2003); M. Sumihama {\em et al.}, Phys. Rev. C {\bf 73}, 035214 (2006).
\bibitem{hallc} G. Niculescu {\em et al.}, Phys. Rev. Lett. {\bf 81}, 1805
     (1998); L. Teodorescu {\em et al.}, Nucl. Phys. {\bf A658}, 362 (1999);
     R.M. Mohring {\em et al.}, Phys. Rev. C {\bf 67}, 052205 (2003).
\bibitem{claspol} D.S. Carman {\em et al.}, Phys. Rev. Lett. {\bf 90}, 131804
     (2003); D.S. Carman {\em et al.}, Phys. Rev. C {\bf 79}, 065205 (2009).
\bibitem{classig} J.W.C. McNabb {\em et al.}, Phys. Rev. C {\bf 69}, 042201 
(2004).
\bibitem{graal} A. Lleres {\em et al.}, Eur.~Phys.~J. A {\bf 31}, 79 (2007);
A. D'Angelo {\em et al.}, {\it ibid} {\bf 31}, 441 (2007); A. Lleres {\em et
al.}, arXiv:0807.3839v1 [nucl-ex].
\bibitem{newsig} R. Bradford {\em et al.}, Phys. Rev. C {\bf 73}, 035202 
     (2006).
\bibitem{newpol} R. Bradford {\em et al.}, Phys. Rev. C {\bf 75}, 035205 
     (2007).
\bibitem{clasep} P. Ambrozewicz {\it et al.},  Phys. Rev. C {\bf 75}, 045203
     (2007). 
\bibitem{clasltp} R. Nasseripour {\it et al.}, Phys. Rev. C {\bf 77}, 065208
     (2008).
\bibitem{abw} R.A. Adelseck, C. Bennhold, and L.E. Wright, Phys. Rev. C
     {\bf 32}, 1681 (1985). 
\bibitem{aws} R. A. Adelseck and L.E. Wright, Phys. Rev. C {\bf 38}, 1965 
     (1988); R.A. Adelseck and B. Saghai, {\em ibid.} {\bf 42}, 108 (1990). 
\bibitem{wjc} Robert A. Williams, Chueng-Ryon Ji, and Stephen R. Cotanch,
     Phys. Rev. C {\bf 46}, 1617 (1992). 
\bibitem{mbh} T. Mart, C. Bennhold, and C.E. Hyde-Wright, Phys. Rev. C 
     {\bf 51}, R1074 (1995); H. Haberzettl, C. Bennhold, T. Mart, and 
     T. Feuster,{\em ibid.} {\bf 58}, R40 (1998); T. Mart and C. Bennhold,
     Nucl. Phys. {\bf A639}, 237c (1998); H. Haberzettl, C. Bennhold, and 
     T. Mart, ACTA Phys. Pol. B {\bf 31}, 2387 (2000); H. Haberzettl, 
     C. Bennhold, and T. Mart, Nucl. Phys. {\bf A684}, 475c (2001). 
\bibitem{mb} T. Mart and C. Bennhold, Phys. Rev. C {\bf 61}, 012201(R) (1999).
\bibitem{korea} M.K. Cheoun, B.S. Han, B.G. Yu, and Il-Tong Cheon, Phys. Rev. C
     {\bf 54}, 1811 (1996); Bong Son Han, Myung Ki Cheoun, K.S. Kim, and
     Il-Tong Cheon, Nucl. Phys. {\bf A691}, 713 (2001). 
\bibitem{sac1} J.C. David, C. Fayard, G.H. Lamot, and B. Saghai, 
     Phys. Rev. C {\bf 53}, 2613 (1996).
\bibitem{sac2} T. Mizutani, C. Fayard, G.-H. Lamot, and B. Saghai, Phys. Rev. C
     {\bf 58}, 75 (1998).
\bibitem{hly} S.S. Hsiao, D.H. Lu, and Shin Nan Yang, Phys. Rev. C. {\bf 61},
     068201 (2000).  
\bibitem{ctls} Wen Tai Chiang, F. Tabakin, T-S.H. Lee, and B. Saghai, 
     Phys. Lett. {\bf B517}, 101 (2001).
\bibitem{ghent} Stijn Janssen, Jan Ryckebusch, Wim Van Nespen, 
     Dimitri Debruyne, and Tim Van Cauteren, Eur. Phys. J. A {\bf 11}, 105 
     (2001); Stijn Janssen, Jan Ryckebusch, Dimitri Debruyne, and 
     Tim Van Cauteren, Phys. Rev. C {\bf 65}, 015201 (2001); Stijn Janssen, 
     Jan Ryckebusch, Dimitri Debruyne, and Tim Van Cauteren, {\em ibid.} 
     {\bf 66}, 035202 (2002); S. Janssen, J. Ryckebusch, and T. Van Cauteren,
     {\em ibid.}, {\bf 67}, 052201(R) (2003); S. Janssen, D.G. Ireland, and 
     J. Ryckebusch, Phys. Lett. {\bf B562}, 51 (2003).
\bibitem{max1} Oren V. Maxwell, Phys. Rev. C {\bf 69}, 034605 (2004); 
\bibitem{max2} Oren V. Maxwell, Phys. Rev. C {\bf 70} 044612 (2004)
\bibitem{max3} Oren V. Maxwell, Phys. Rev. C {\bf 76}, 014621 (2007).
\bibitem{us} A. Usov and O. Scholter, Phys. Rev. C {\bf 72}, 025205 (2005). 
\bibitem{bonn} A.V. Sarantsev, V.A. Nikonov, A.V. Anisovich, E. Klempt, and 
     U. Thoma, Eur. Phys. J. A {\bf 25}, 441 (2005);  A.V. Anisovich,
     V. Kleber, E. Klempt, V.A. Nikonov, A.V. Sarantsev, and U. Thoma,
     Eur. Phys. J. A {\bf 34}, 243 (2007);  V.A. Nikonov,
     A.V. Anisovich, E. Klempt,  A.V. Sarantsev, and U. Thoma,
     Phys. Lett. B {\bf 662}, 245 (2008).
\bibitem{kaiser} N. Kaiser, T. Waas, and W. Weise, Nucl. Phys. A {\bf
     612}, 297 (1997).
\bibitem{feuster} T. Feuster and U. Mosel, Phys. Rev. C {\bf 59}, 460
     (1999).
\bibitem{jdiaz} B. Julia-Diaz, B. Saghai, T. S. Lee, and F. Tabakin,
     Phys. Rev. C {\bf 73}, 055204 (2006).
\bibitem{bydzovsky} See, for example, the discussion in T. Mart and
A. Sulaksono, Phys. Rev. C {\bf 74}, 055203 (2006); and P. Byd\v
     zovsky' and T. Mart, Phys. Rev. C {\bf 76}, 065202 (2007).
\bibitem{muk} M. Benmerrouche, R.M. Davidson, and Nimai C. Mukhopadhyay, 
     Phys. Rev. C {\bf 39}, 2339 (1989).
\bibitem{pdt} W.M. Yao et al., J. Phys. G {\bf 33}, 1 (2006).
\bibitem{quark} Simon Capstick and W. Roberts, Phys. Rev. D {\bf 58}, 074011 
     (1998).
\bibitem{klein} F.J. Klein {\it et al.}, Jefferson Lab Experiment E02-112.
\bibitem{dumb} O.Dumbrajs, R. Koch, H. Pilkuhn, G.C. Oades, H. Behrens, 
      J.J. de Swart, and P. Kroll, Nucl. Phys. {\bf B216}, 277 (1983).
\bibitem{deswart} J.J. deSwart, Rev. Modern Phys. {\bf 35}, 916 (1963).
\bibitem{gencot} Ignacio J. General and Stephen R. Cotanch, Phys. Rev. C 
{\bf 69},035202 (2004).

\end{thebibliography}
\end{document}